\documentclass{aa-nolinenum}
\usepackage[varg]{txfonts}
\usepackage{xspace}
\usepackage{upgreek}
\usepackage[breaklinks=true, hidelinks]{hyperref}

\newcommand{\mum}{\textmu{}m\xspace}

\makeatletter
\renewcommand*\aa@pageof{, page \thepage{} of \pageref*{LastPage}}
\makeatother

\usepackage{orcidlink}
\usepackage{soul}
\usepackage{xcolor}
\definecolor{darkorange}{rgb}{1.0, 0.55, 0.0}

\newcommand{\revision}[1]{#1}

\begin{document}

\title{PDRs4All\\ XVI. Tracing aromatic infrared band characteristics in photodissociation region spectra with PAHFIT in the JWST era}

\author{%
    Dries Van De Putte\inst{\ref{WOntarioPA} \orcidlink{0000-0002-5895-8268}} \and
    Els Peeters \inst{\ref{WOntarioPA}, \ref{WOntarioIESE}, \ref{CarlSagan} \orcidlink{0000-0002-2541-1602}} \and
    Karl D. Gordon \inst{\ref{STScI}, \ref{Ghent} \orcidlink{0000-0001-5340-6774}} \and 
    J. D. T. Smith \inst{\ref{Toledo} \orcidlink{0000-0003-1545-5078}} \and
    Thomas S.-Y. Lai \inst{\ref{IPAC} \orcidlink{0000-0001-8490-6632}}\and
    Alexandros Maragkoudakis \inst{\ref{ARC}} \and
    Bethany Schefter \inst{\ref{WOntarioPA}, \ref{WOntarioIESE}} \and
    Ameek Sidhu \inst{\ref{WOntarioPA}, \ref{WOntarioIESE}}\and 
    Dhruvil Doshi \inst{\ref{WOntarioPA}} \and
    Olivier Berné \inst{\ref{ToulouseIRAP}} \and
    Jan Cami \inst{\ref{WOntarioPA}, \ref{WOntarioIESE}, \ref{CarlSagan} \orcidlink{0000-0002-2666-9234}} \and
    Christiaan Boersma \inst{\ref{ARC} \orcidlink{0000-0002-4836-217X}} \and
    Emmanuel Dartois \inst{\ref{ISMO} \orcidlink{0000-0003-1197-7143}} \and
    Emilie Habart \inst{\ref{SaclayIAS}} \and
    Takashi Onaka \inst{\ref{TokyoAstro} 
    \orcidlink{0000-0002-8234-6747}} \and
    Alexander G.~G.~M. Tielens \inst{\ref{Leiden}, \ref{Maryland}} }

\institute{
  Department of Physics \& Astronomy, The University of Western Ontario, London ON N6A 3K7, Canada
  \label{WOntarioPA}     \and 
  Institute for Earth and Space Exploration, The University of Western Ontario, London ON N6A 3K7, Canada
  \label{WOntarioIESE}          \and 
  Carl Sagan Center, SETI Institute, 339 Bernardo Avenue, Suite 200, Mountain View, CA 94043, USA
  \label{CarlSagan}         \and
  Space Telescope Science Institute, 3700 San Martin Drive, Baltimore, MD 21218, USA
  \label{STScI} \and
  Sterrenkundig Observatorium, Universiteit Gent, Gent, Belgium
  \label{Ghent} \and 
  Ritter Astrophysical Research Center, University of Toledo, Toledo, OH 43606, USA
  \label{Toledo} \and
   IPAC, California Institute of Technology, 1200 E. California Blvd., Pasadena, CA 91125
  \label{IPAC} \and
  NASA Ames Research Center, MS 245-6, Moffett Field, CA 94035-1000, USA
  \label{ARC} \and
  Institut de Recherche en Astrophysique et Planétologie, Université Toulouse III - Paul Sabatier, CNRS, CNES, 9 Av. du colonel Roche, 31028 Toulouse Cedex 04, France
  \label{ToulouseIRAP}         \and 
  Institut des Sciences Moléculaires d'Orsay, CNRS, Universit\'e Paris-Saclay, Orsay, France
  \label{ISMO} \and
  Institut d'Astrophysique Spatiale, Université Paris-Saclay, CNRS, Bâtiment 121, 91405 Orsay Cedex, France
  \label{SaclayIAS}        \and 
 Department of Astronomy, Graduate School of Science, The University of Tokyo, 7-3-1 Bunkyo-ku, Tokyo 113-0033, Japan
  \label{TokyoAstro} \and
   Leiden Observatory, Leiden University, P.O. Box 9513, 2300 RA Leiden, The Netherlands
  \label{Leiden} \and
  Astronomy Department, University of Maryland, College Park, MD 20742, USA
  \label{Maryland}
}

\date{Accepted June 26, 2025}

\abstract
{Photodissociation regions (PDRs) exhibit strong emission bands between 3-20 \mum known as the aromatic infrared bands (AIBs), and they originate from small carbonaceous species such as polycyclic aromatic hydrocarbons (PAHs) excited by UV radiation.
The AIB spectra observed in Galactic PDRs are considered a local analog for those seen in extragalactic star-forming regions.
Recently, the PDRs4All JWST program observed the Orion Bar PDR, revealing the subcomponents and profile variations of the AIBs in very high detail.
}
{
We present the Python version of PAHFIT, a spectral decomposition tool that separates the contributions by AIB subcomponents, thermal dust emission, gas lines, stellar light, and dust extinction.
We aim to provide a configuration that enables highly detailed decompositions of JWST spectra of PDRs ({3.1-26~\mum}) and to test if the same configuration is suitable to characterize AIB emission in extragalactic star forming regions.
}
{
We determined the central wavelength and FWHM of the AIB subcomponents by fitting selected segments of the Orion Bar spectra and compiled them into a ``PDR pack'' for PAHFIT.
We tested the PDR pack by applying PAHFIT to the full {3.1-26}~\mum PDRs4All templates.
We applied PAHFIT with this PDR pack and the default continuum model to seven spectra extracted from the central star forming ring of the galaxy NGC7469.
}
{
We introduce an alternate dust continuum model to fit the Orion Bar spectra, as the default PAHFIT continuum model mismatches the intensity at 15-26~\mum.
\revision{Using the PDR pack and the alternate continuum model, PAHFIT} reproduces the Orion Bar template spectra with residuals of a few percent.
A similar performance is achieved when applying the PDR pack to the NGC7469 spectra.
We provide PAHFIT-based diagnostics that trace the profile variations of the 3.3, 3.4, 5.7, 6.2, and 7.7~\mum AIBs and thus the photochemical evolution of the AIB carriers. 
The 5.7~\mum AIB emission originates from at least two subpopulations, one more prominent in highly irradiated environments and one preferring more shielded environments.
{Smaller PAHs as well as very small grains or PAH clusters both} thrive in the more shielded environments of the molecular zone in the Orion Bar.
Based on these new diagnostics, we \revision{show and} quantify the \revision{strong similarity of} the AIB profiles observed in NGC7469 to the Orion Bar template spectra.
}
{}


\authorrunning{Van De Putte et al.}
\titlerunning{PDRs4All and PAHFIT}
\maketitle

\section{Introduction}

A key component of the infrared (IR) emission from the interstellar medium of galaxies is the near-IR (NIR) and mid-IR (MIR) emission by complexes of broad features spanning 3-20~\mum.
In this series of papers (PDRs4All), these are referred to as the aromatic infrared bands (AIBs).
In the literature, they are also referred to as ``PAH emission'' or ``PAH bands,'' owing to the interpretation that they originate from vibrational transitions of polycyclic aromatic hydrocarbons (PAHs) under the influence of pumping by far-ultraviolet (FUV) photons \citep{Leger1984, Allamandola1985}.
The main AIBs peak at 3.3, 5.2, 6.2, 7.7, 8.6, 11.3, 12.7, 16.4, and 17.4~\mum, and many of them appear to consist of two or more subcomponents. 
Previous observations have revealed profile variations of the bands depending on the environment, for example, as seen in Infrared Space Observatory (ISO) spectra \citep{Peeters2004}.
The observed variations can be combined with knowledge from molecular modeling and lab experiments to learn about the photochemical evolution of the carriers and how they are affected by local physical conditions such as the radiation field or the temperature, density, and chemical composition of the gas \citep[and references therein]{Tielens2008}.

The AIBs are exceptionally bright in photodissociation regions (PDRs), which typically manifest where parts of molecular clouds are irradiated by FUV photons from young stars formed within them \citep{Tielens1985}.
PDRs are also found in the diffuse interstellar medium (ISM) \citep{Wolfire2003}, and most of the atomic and molecular gas in the ISM of galaxies are in regions under PDR conditions \citep[and references therein]{Wolfire2022}. 
A large fraction of the IR emission of galaxies originates from PDRs, particularly for (ultra)luminous IR galaxies.
At the typical spatial scales of around 1 kpc, the observed spectra are a luminosity weighted average across an ensemble of PDRs and local environments.
It is therefore expected that the AIB spectra observed in extragalactic observations are well represented by exposed regions, such as PDRs, found in our own Galaxy \citep{Maragkoudakis2018}.
With the integral field unit (IFU) modes of the NIRSpec and MIRI instruments \citep{2022Jakobsen, 2022Boker, 2023Wright} on board the James Webb Space Telescope \citep[JWST;][]{jwst}, the spectral variations of the AIB profiles can be spatially resolved together with the HI to H$_2$ transition front located in PDRs.
The AIB variations can be studied in the context of the local physical conditions by combining those observations with the many available gas-line diagnostics that trace the physical and chemical conditions in the PDR \citep[e.g., those included in the PDR Toolbox,][]{Pound2023}.

The ``PDRs4All'' program for JWST \citep[ERS ID 1288]{Berne2022} produced spectroscopic observations of the AIB emission in the Orion Bar PDR with an unprecedented combination of depth (S/N) and spatial resolution in both imaging and integral field spectroscopy \citep{Habart2024, Peeters2024}.
The overview of the AIB contents by \citet{Chown2024} has revealed the richness in subcomponents of the AIB features by comparing the spectra extracted from different regions in the Orion Bar, and they list approximate central wavelengths as well as likely identifications in terms of vibrational modes. 
In terms of the classification scheme introduced by \citet{Peeters2002} and \citet{vanDiedenhoven2004}, these data mainly reveal variations within class A, which encompasses objects such as HII regions, PDRs, reflection nebulae, and galaxies.
There are substantial variations in the inter-band ratios and relatively subtle profile variations within each band.
Examples are the broadening of the 5.2, 6.2, 7.7, and 12.7~\mum features as well as the central wavelength shift and the asymmetry caused by the broad red wing of the 11.2~\mum feature.

To quantify the AIB profile variations and inter-band ratios, the partially blended emission features have to be separated from each other and from other emission components such as gas lines, the stellar continuum, and the thermal dust continuum.
All of those components can also be affected by dust attenuation and absorption by ices \citep{Lai2024}. 
For this task, a class of methods known as ``spectral decomposition'' can be applied in order to provide consistent feature strengths that can be used for diagnostic investigations.
One such tool is PAHFIT \citep{Smith2007}, and to this day it has mainly been applied to extract AIB strengths from Spitzer Infrared Spectrograph \citep[IRS;][]{Houck2004} observations, for which it was originally designed \citep[e.g.,][]{Gordon2008, Sellgren2010, Treyer2010, Stock2013, Hemachandra2015, Smercina2018, Maragkoudakis2025}.
Other quantities of interest simultaneously determined by PAHFIT are the strength of gas lines \citep[e.g.,][]{Tarantino2024} and the depth of the 10 and 20~\mum silicate features due to dust attenuation \citep[e.g.,][]{Goulding2012}.

The original parameters used for the Spitzer data have since been modified to allow for analysis of observations from other observatories such as ISO \citep{Maragkoudakis2018} or AKARI \citep{Lai2020}.
Decomposing the spectra observed by JWST poses new challenges due to the higher resolution ($R \sim 1000$-3000), as PAHFIT has mainly been used at $R \sim 100$.
In this work, {we} introduce the Python version of PAHFIT and new features and concepts that were implemented in the context of JWST spectra.
We aim to update the model components (AIBs and continuum) to fit the Orion Bar spectra of PDRs4All, which will provide a PAHFIT setup suitable for other PDR spectra in the JWST era (the ``PDR pack'').
To explore its broader applicability, we applied this PDR pack to spectra of the central starburst ring in the galaxy NGC7469 and used the results to study the similarity of the AIBs in this star forming galaxy to those in the Orion Bar.
Given the high spectral resolution and signal-to-noise of the JWST data, we also explored new diagnostics that probe variations in the AIB profiles that are driven by the photochemical evolution of the PAH population. 
 
We note that adapting PAHFIT for JWST spectroscopy of PDRs is one of the science-enabling products by the PDRs4All collaboration. Hence, this paper is released as a part of the PDRs4All series. 
This work is organized as follows:
In Sect.~\ref{sec:data}, we describe the Orion Bar and NGC7469 spectra.
In Sect.~\ref{sec:method}, we introduce the new version of PAHFIT and explain the strategy to adjust the settings of the spectral decomposition based on the Orion Bar spectra.
We then apply the resulting PDR pack and validate the fitting performance in Sect.~\ref{sec:results}.
In the discussion, we introduce diagnostics that quantify the variations of the AIB profiles at 3.3, 5.7, 6.2, and 7.7~\mum, and we investigate their power in probing the photochemical evolution of the AIB carriers. 
We employed these diagnostics to reveal the similarities between the NGC7469 spectra and the Orion Bar template spectra.
We conclude with a summary in Sect.~\ref{sec:conclusions}.

\section{Data}
\label{sec:data}

\subsection{Orion Bar PDR spectra} 

We aim to test and fine-tune PAHFIT using the JWST PDRs4All data of the Orion Bar, as these data offer high S/N spectra over the complete range of AIBs.
We use the PDRs4All ``templates'' \citep{Chown2024}, which are five spectra extracted from a pre-defined set of rectangular apertures on the sky, chosen to represent and sample emblematic regions of the PDR, as defined by \citet{Peeters2024}.
We refer to these templates by their names: ``HII'' represents the HII region and a face-on PDR in the background; ``Atomic PDR'' is  located in the neutral region just behind the ionization front and is the brightest region for most AIBs; ``DF1,'' ``DF2,'' and ``DF3'' represent the three dissociation fronts present in the field of view.

The template spectra extracted using the above apertures are periodically updated and delivered to the public as part of PDRs4All, and the updated templates used in this work will be available electronically on the PDRs4All website (PDRs4All.org) and via the CDS. 
We extracted the spectra from updated JWST data cubes, produced with more recent versions of the JWST pipeline and the CRDS references files (JWST 1.15.1, CRDS context jwst\_1253.pmap for NIRSpec and jwst\_1276.pmap for MIRI).
After the aperture extractions from the individual data cubes, additive offsets were applied to even out (`stitch') flux mismatches in the wavelength overlap regions of the spectral segments.
The stitching method uses MRS channel 1 SHORT as the reference ($\lambda=4.90-5.74$~\mum) and the additive offsets are calculated by taking the difference of the medians of the flux in the regions of overlap, resulting in offsets of the order 1-3\% \citep{VanDePutte2024}.
Some of the JWST pipeline wrapper scripts and post-processing scripts used for this reprocessing are also available in a public repository.\footnote{\url{https://github.com/PDRs4All/PDRs4All}}

\subsection{NGC7469 star forming ring spectra}

We required observations of the AIB emission in a galaxy with similar depth as the Orion Bar spectra, and for this purpose, we selected publicly available JWST data of the galaxy NGC7469 observed with JWST as part of the GOALS program (ERS ID 1328, \citealt{Armus2023}). 
This target is preferable over other candidates in the GOALS sample due to its lower dust attenuation.
We downloaded the uncalibrated data from MAST for the NIRSpec F290LP and MIRI MRS observations, and employed our PDRs4All pipeline script and stitching method to reduce these data as was done for the Orion Bar (pipeline 1.17.1, jwst\_1322.pmap).
We also reduced the dedicated background observations for the MIRI data, and used them to perform the master\_background step of the {stage 3} pipeline.
We note that for one aperture (SF1 defined below), the NIRSpec flux was roughly 1.8 times higher than that of the MRS channel 1 in the region of overlap.
This is the reason why we preferred additive offsets for matching the spectral segments, as opposed to multiplicative ones.
For the other apertures and the regions of overlap of the MRS segments, the mismatch was only a few percent, similar to that documented by \citet{VanDePutte2024}.

\begin{figure}[tbh]
    \centering
    \includegraphics[]{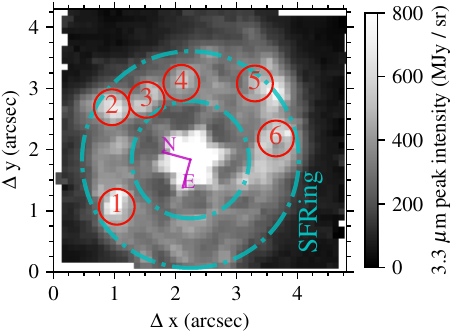}
    \caption{Apertures used for extracting NGC7469 spectra from NIRSpec and MIRI IFU data cubes.
    The background image is a single NIRSpec slice near the peak of the 3.3 \mum band ($3.343$ \mum due to redshift), and the north and east indicators at the center indicate the orientation.
    The red circles (radius of 0.3'') represent the SF1 to SF6 apertures as numbered.
    The cyan circles indicate the circular annulus region used to extract the average spectrum of the SFRing surrounding the AGN.
    \label{fig:apertures}}
\end{figure}

\begin{figure*}[tbh]
\centering
\includegraphics[scale=0.9]{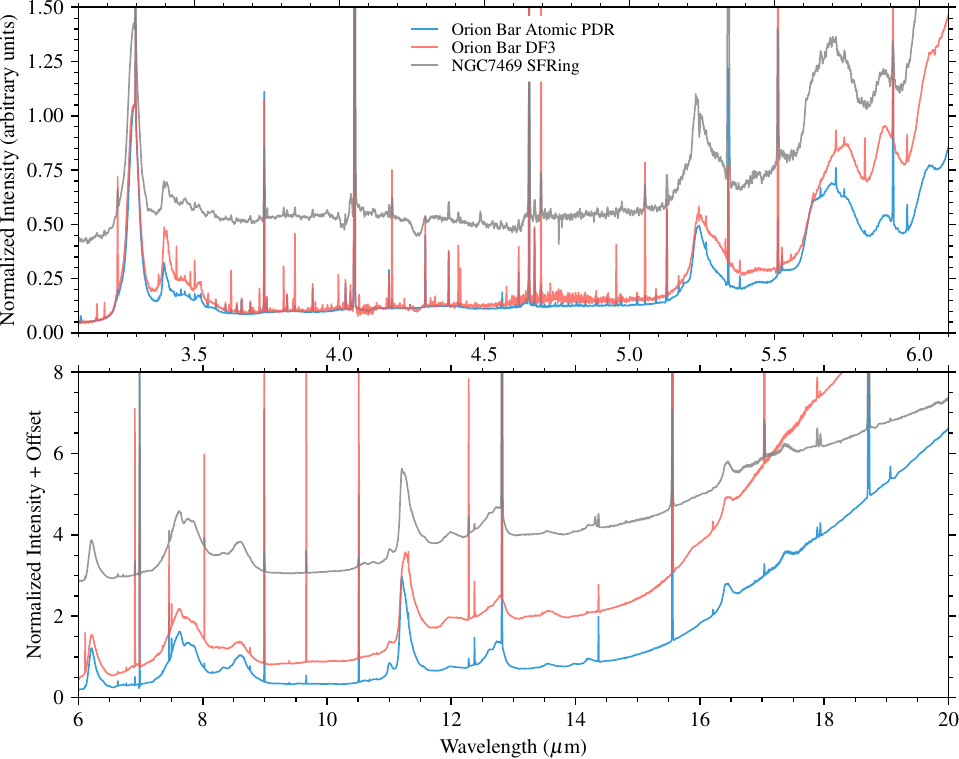}
\caption{Overview of a selection of Orion Bar and NGC7469 spectra used in this work.
The spectra {($F_\nu$ units, MJy sr$^{-1}$)} were scaled to the amplitude of the bands at 3.3~\mum (upper panel) and 6.2~\mum (lower panel).
The NGC7469 SFRing spectrum was redshift corrected ($z = 0.016317$), and for the lower panel, a vertical offset was applied.}
\label{fig:spectra}

\end{figure*}

NGC7469 exhibits an active galactic nucleus (AGN) surrounded by a bright circumnuclear ring of star formation \citep{Lai2022, Armus2023}.
We selected six circular apertures on the star-forming ring with a radius of 0.3 arcsec, shown in Fig.~\ref{fig:apertures}, from which spectra were extracted using aperture photometry. 
The apertures labeled SF1-SF6 were selected in approximately the same way as \citet[][first column of their Fig.~6]{Donnan2024}.
We also extracted the average spectrum of the entire star-forming ring (hereafter ``SFRing''), using an annulus centered at 345.8151484d +8.8739239d, with inner and outer radii of 0.98'' and 1.81'' respectively.
At first sight, the SFRing spectrum looks very similar to the Orion Bar PDR spectra (Fig.~\ref{fig:spectra}).
Before using them in the figures or in the analysis, the wavelengths of the NGC7469 spectra were shifted to the rest frame assuming a constant $z =  0.0163$.
We note that the AGN emission exhibits very different spectra compared to the starburst ring \citep{Lai2023}, and this regime was therefore excluded.

\subsection{Additional data preparation for fitting}
\label{sec:cleandata}

A selection of spectra used in this work, as extracted from the JWST products, is shown in the overview of Fig.~\ref{fig:spectra}, where a linear scale is used to emphasize the difference in the slope of the continuum between NGC7469 and the Orion Bar.
A number of additional preparation steps were applied prior to using the spectra as input for PAHFIT.
Firstly, a few additional wavelength intervals were removed from the data.
The 3.98-4.17 \mum interval was removed to avoid instrumental artifacts, as the NIRSpec chip gap is located here for the G395H mode.
The 4.54-4.85 \mum interval was removed because a combination of a deuterated PAH feature and many molecular emission lines appear here {\citep{Peeters2024, Draine2025}}.
Lastly, the 6.77-7.02 \mum interval was removed to avoid artifacts nearby the bright [\ion{Ar}{ii}] 6.99 \mum line \citep{VanDePutte2024}.

{We note that the angular resolution of the spectra depends on the wavelength.
The coarsest spaxels (MRS channel 4) have a size of around 0.35''.
In this work, we therefore limit the analysis to a number of regions which are averaged over several spatial elements (spaxels), and the smallest regions we use are the circular apertures in Fig.~\ref{fig:apertures}, with a radius of 0.3''. 
For analyses that apply to individual spaxels, a resolution matching based on the point spread function would be the best approach, where the resolution of the shorter wavelength maps is degraded to that of MRS channel 4. 
}

{Taking averages over multiple spaxels }exacerbates the fact that not all uncertainties are tracked by the JWST pipeline, resulting in unrealistically small uncertainties (S/N $\gtrsim 1000$).
Even with the detailed tuning represented in this work, the model cannot reproduce the data within those very small uncertainties.
We find this makes the PAHFIT fitting algorithm (see Sect.~\ref{sec:pahfit}) more susceptible {to} overfitting or local minima.
Considering the known calibration uncertainties of MIRI MRS and NIRSpec, having a S/N $> 200$ is almost certainly unrealistic, and we replaced the uncertainties on the data by applying a S/N limit of 150 to all spectra.

\section{Method}
\label{sec:method}

\subsection{PAHFIT development}
\label{sec:pahfit}

The original concept of PAHFIT was introduced by \citet{Smith2007}, with the main code written in IDL.
As part of the PDRs4All program, development was started for a Python port of PAHFIT\footnote{\url{https://github.com/pahfit/pahfit/}}, which is delivered as a science-enabling product to the community (SEP5\footnote{\url{https://pdrs4all.org/seps/}}).
Hereafter, we refer to these versions as ``IDL PAHFIT'' and ``Python PAHFIT,'' respectively, where disambiguation is necessary.
Below, we provide a brief summary of the PAHFIT concept based on the IDL implementation, and follow with a description of the improved features offered by the Python port.

\subsubsection{PAHFIT concept and IDL implementation}

The core functionality of PAHFIT is its multi-component model, consisting of ``dust feature'' components that represent the AIBs, a combination of ``stellar'' and ``dust continuum'' components, and ``line'' components that can be used to take into account gas lines that blend with the AIBs. 
These additive emission components can be combined with a multiplicative component of the model to include an attenuation curve or individual absorption features \citep[Eq.~1]{Smith2007}.
Each component has a number of parameters depending on its type, each of which can be varied or kept constant.
The fitting of these parameters occurs via a least squares minimization using the Levenberg-Marquardt algorithm.

A dust feature in IDL PAHFIT is represented by a Drude profile with a certain amplitude $b_r$, central wavelength $\lambda_r$, and width $\gamma_r$ (fractional FWHM).
The motivation for this choice is given by \citet{Smith2007}, and we reproduce the analytical equation for the profile here:
\begin{equation}
    I^{(r)}_\nu(\lambda) = \frac{b_r \gamma_r^2}{(\lambda/\lambda_r - \lambda_r/\lambda)^2 + \gamma_r^2}.
    \label{eq:drudeamplitude}
\end{equation}
By default, $\lambda_r$ and $\gamma_r$ are kept constant during the fit, but they can be configured to vary if needed.
We note that the central wavelength definitions for both lines dust emission features modeled by PAHFIT are defined in the rest frame, and therefore the assumption is that the input spectra have also been shifted into the rest frame.

The shape of the continuum model is crucial for obtaining reliable fits of all other features.
The PAHFIT continuum model represents dust populations with different equilibrium temperatures using multiple ``modified blackbody'' (MBB) components, of which the intensity is given by
\begin{equation}
I^{(r)}_\nu(\lambda) = B_\nu(\lambda; T_r) (\lambda_0 / \lambda)^2,    
\label{eq:mbb}
\end{equation}
where $B_\nu(\lambda; T_r)$ is the blackbody intensity (Planck's law) evaluated at a temperature $T_r$, and $\lambda_0$ is an arbitrary reference wavelength ($\lambda_0 = 9.7$~\mum).
In IDL PAHFIT, the standard dust continuum configuration consists of eight MBBs, with temperatures of 35, 40, 50, 65, 90, 135, 200, and 300 K, although it is not uncommon to adjust these temperatures or add components representing warm dust (500-1200 K), in AGN environments for example.
On the blue end of the spectrum, a regular blackbody function $B_\nu(\lambda; T)$ with $T = 5000$ K is added to model the red tail of the stellar continuum.

The dust attenuation model in PAHFIT is applied as a multiplier $f$ affecting the total spectrum, which is the sum of all other components: $ I_\nu(\lambda) = f(\lambda) \sum_r I^{(r)}(\lambda)$.
The geometry of this model is ``fully mixed'' by default, which assumes that the dust and the emitting sources (stars, gas, excited PAHs) are equally distributed. 
Alternatively, a screen geometry can be applied.
Mathematically, the two options read
\begin{gather}
    f_\text{mixed}(\lambda) = \frac{1 - e^{-\tau_\lambda}}{\tau_\lambda}\\
    f_\text{screen}(\lambda) = e^{-\tau_\lambda},
\end{gather}
    where $\tau_\lambda$ is an optical depth model consisting of a power law with two silicate features added \citep{Smith2007}.
The free parameter for the attenuation is $\tau_{9.7}$, and the optical depth at the peak of the silicate feature is at 9.7 \mum.

\subsubsection{Python port and new features}
\label{sec:pythonpahfit}

The Python port of PAHFIT implements the same concepts and types of components as above, with a few key differences.
The parameterization of the dust features was changed so that the power of each Drude profile is preserved when the FWHM is varied.
The intensity of a Drude profile is now derived from its total power $P_r$ (e.g., W m$^{-2}$ sr$^{-1}$) and width, rather than the amplitude $b_r$ (e.g., MJy / sr), while the width is defined in wavelength units FWHM$_r$ instead of the fractional FWHM $\gamma_r$.
The implementation uses Eq.~\ref{eq:drudeamplitude} with the following substitutions:
\begin{gather}
    \gamma_r = \text{FWHM}_r / \lambda_r\\
    b_r = \frac{2 P_r \lambda_r}{\pi c \gamma_r} f_\text{unit}\\
    f_\text{unit} = \frac{u_\text{power} \mu \text{m}}{\text{MJy sr}^{-1}}
    \label{eq:drudepower},
\end{gather}
where the power unit $u_\text{power}$ is $10^{-22}$ Wm$^{-2}$.
The units in the above equations were chosen so that the order of magnitude of $P_r$ is between $10^0$ and $10^5$ in order to avoid numerical issues that arise with very large or small numbers.

Python PAHFIT introduces a new system for configuring the model via ``science packs.''
A science pack is a list of features and their parameters to be included in the fit, and a different list may be needed depending on the type of target (e.g.,~PDRs, AGNs, highly obscured galaxies).
It is therefore a future goal to provide a collection of science packs, which contain reasonable default configurations for fitting spectra of several key classes of astronomical objects.
The science packs for Python PAHFIT are written in the YAML data serialization language, and offer various conveniences and shorthand notations to specify parameter values and bounds, making them straightforward to customize (see PAHFIT documentation\footnote{\url{pahfit.readthedocs.io}}).

Currently, one science pack is included in the PAHFIT Python package and it is called the ``classic'' pack.
It was designed to have the same contents as the IDL PAHFIT v1.2 model \citep{Smith2007}, which was optimized for Spitzer IRS spectra.
This means that the classic pack has equivalent FWHM and central wavelengths for the AIBs, and the same default set of temperatures for the continuum.
We note that there is already published work that makes use of Python PAHFIT and the classic pack \citep[e.g.,][]{Maragkoudakis2022, Maragkoudakis2025}. 
This paper documents the first release of a dedicated science pack for use with JWST data, hereafter called the ``PDR pack.''

Python PAHFIT also introduces ``instrument packs,'' which contain resolution curve models for various observatories, instruments, and spectroscopic configurations.
When unresolved emission lines are observed, or when the spectral resolution is low compared to the width of the narrowest AIB subcomponents, the FWHM of the affected features needs to be adjusted accordingly.
Currently, this concept is applied to the gas lines, where the FWHM of each line is derived from the resolution curve \citep[e.g.,][for MIRI MRS]{Labiano2021}.
The instrumental broadening of narrow dust emission features will be implemented in the future, but this effect is only relevant in the low spectral resolution case ($\sim100$), and is minimal for the JWST data presented here.
This contrasts with IDL PAHFIT, were the FWHM were determined empirically from spectra with $R \sim 100-150$, and as such this slight broadening was encoded in the AIB parameters, while the FWHM of the gas lines had to be adjusted by the user depending on the spectral resolution.

By introducing the science pack and instrument pack concepts, the physical and instrumental parts of the model are now separated.
For each type of spectrum (type of object observed) and each instrument (spectral resolution), one science pack and one instrument pack can be combined to set up a suitable model to fit the observations.
The use case for this is that a single science pack can be reused for different instruments, considering that there are now multiple observatories and archives from which NIR or MIR spectra are available; instrument packs have been set up for Spitzer, ISO, and JWST and will be developed for AKARI.

\subsection{Science pack for JWST spectra of PDRs}

\subsubsection{Alternate continuum tuning and model}
\label{sec:pahfitcontinuummodel}

\begin{figure}[tb]
    \centering
    \includegraphics[scale=0.9]{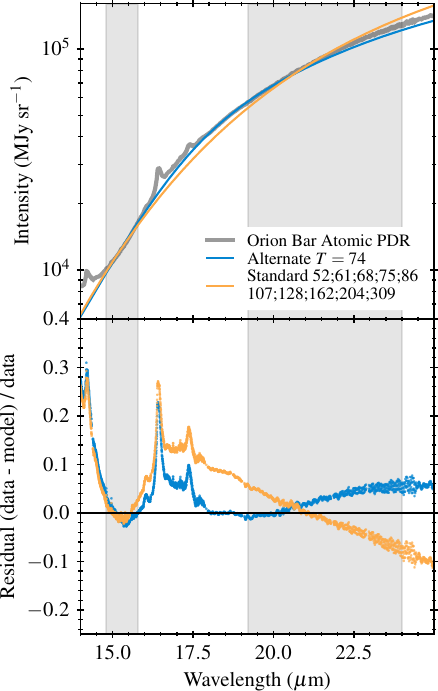}
    \caption{Demonstration of how the default modified blackbody fails to reproduce the Orion Bar continuum.
    Multiple standard MBB components (orange) and a single-component alternate MBB (blue) were fit to two specific sections of the 
    Orion Bar atomic PDR spectrum (shaded areas).
    The legend indicates the temperatures (K) of the components.
    Attenuation was included and yields $\tau_{9.7} = 0.06$ for the standard MBB model, zero for the alternate model.
    The standard MBB cannot simultaneously reproduce the intensities and slopes near 15 and 22 \mum, undercutting the 16-18 \mum complex by 10\%.
    The alternate MBB produces smaller residuals with just a single 74 K component.}
    \label{fig:defaultvsspecial}
\end{figure}

We find that the standard PAHFIT continuum components (Eq.~\ref{eq:mbb}) cannot simultaneously fit the $\sim 15$~\mum and 18-26~\mum ranges of the steep continuum observed in the Orion Bar.
In Fig.~\ref{fig:defaultvsspecial} we illustrate this by fitting a model consisting of many MBB components to the Orion Bar atomic PDR spectrum, while restricting the fit to the 14.5-15.5 and 19-24~\mum ranges.
For this demonstration, the resulting stellar continuum and dust attenuation are negligible.
The fit weights used are $\sigma^{-2}$ with $\sigma$ equal to 1\% of the intensity, which emphasizes the 15~\mum region.
When the continuum level near 15~\mum is well approximated, the resulting slope at longer wavelengths mismatches that of the data.
For all Orion Bar templates, this prevents us from obtaining reliable fits for the prominent 16.4 and 17.4~\mum AIB features, as the resulting continuum level underneath these features is too low (see residuals in Fig.~\ref{fig:defaultvsspecial}).

We note that similar issues with the continuum were identified in other nebulae (HII regions) with similarly steep dust continua \citep[e.g.,][]{Maragkoudakis2018} and additional dust features were included in the PAHFIT model as a solution.
\citet{Foschino2019} apply an extinction factor to the standard MBB shapes, and successfully fit ISO-SWS spectra of the Orion Bar, though their study is limited to the 2.6-15~\mum range.
Others compute a dust emissivity curve based on grain models, taking into account grains with different sizes, temperatures, and optical properties \citep[e.g.,][]{Marshall2007}.
A population of grains with a given equilibrium temperature $T_a$ has an emissivity proportional to $I_\nu(\lambda) = B_\nu(\lambda; T) C_\text{abs}(\lambda, a)$, where $C_\text{abs}$ is the absorption cross section, according to Kirchoff's law of thermal radiation.
Based on this, we set up the partially ad hoc prescription 
\begin{equation}
I^{(r)}_\nu(\lambda) = B_\nu(\lambda; T) \frac{A(\lambda)}{A(V)}
\label{eq:altmbb},
\end{equation}
{where $A(\lambda) / A(V)$ is an extinction curve.}
In other words, we replaced the $\lambda^{-2}$ term of the standard MBB (Eq.~\ref{eq:mbb}) by an extinction curve.
The need to replace the $\lambda^{-2}$ power law for certain sources is expected, as the emissivity index (the power law exponent $\beta$) is known to vary \citep{Hildebrand1983}.
The spectral index originates from an approximation for the dust opacity factor in the equation of the flux density of an isothermal source \citep[e.g.,\ Eq.~1 of][]{Shetty2009}.
As extinction curves are direct probes of an averaged opacity over a dust population, substituting $\lambda^{-2}$ by $A(\lambda)$ can be considered an alternative but conceptually similar approximation.

To apply Eq.~\ref{eq:altmbb}, {we compute the extinction $A(\lambda)$ using the ``G23'' class from the publicly available ``dust\_extinction'' Python package \citep{Gordon2024}.
This is an $R_V$-dependent average of the extinction observed in the Milky Way, parameterized as $A_\text{G23}(\lambda; R_V)$ \citep{Gordon2023}, based on measurements across the FUV to MIR range \citep{Gordon2009, Fitzpatrick2019, Gordon2021, Decleir2022}.
We chose a value of $R_V = 5.5$ as it is considered suitable for the dust in the Orion Bar and consistent with \citet{Peeters2024}.}
The silicate extinction features in this curve introduce bumps at 10 and 20~\mum in the MBB curve, and the latter bump increases the emission at 20~\mum relative to that at 15~\mum, which is what we need for the Orion Bar spectrum considering Fig.~\ref{fig:defaultvsspecial}.
The comparison in Fig.~\ref{fig:defaultvsspecial} shows that just a single temperature reproduces the slope change between the 15~\mum and 22~\mum continuum, and is more suitable for extracting the 15-18~\mum AIBs.
Qualitatively, this shape is similar to the dust continuum components by \citet[][see their Fig.~7]{Marshall2007}, which are based on emissivity models for silicate grains.
The presence of silicate absorption in the Orion Bar is likely, as the H$_2$ 0-0 S(3) 9.6649 \mum line intensity is lower than expected, when compared to a value interpolated between the adjacent S(2) and S(4) lines in a rotational diagram \citep{VanDePutte2024}. 
Silicate emission is also observed in the d203-504 disk located in the Orion Bar field of view, though this is a point-like source (Schroetter et al., in prep.).

A more general and improved version of the alternate MBB will be included in the official PAHFIT release upon publication of this work. 
The version used in this work is implemented in an experimental PAHFIT branch dedicated to this work, of which a snapshot will provided as a separate release on the PAHFIT GitHub page.
As with the standard continuum, we set up a list of fixed MBB temperatures to define the continuum model, and included the following 13 temperatures: 45, 60, 75, 90, 105, 120, 160, 200, 240, 280, 320, 360, and 400 K.
As is the case for the classic set of continuum temperatures, the individual components are ascribed no special significance, and are designed to span a range of cool to warm emission spanning 5-25~\mum.
Caveats related to the continuum are summarized in Sect.~\ref{sec:caveatcontinuum}.
\subsubsection{Dust feature tuning strategy}
\label{sec:tuningstrategy}

In the PAHFIT model, individual AIBs are assumed to consist of one or more subcomponents (``dust features'') described by Drude profiles.
Tuning the dust features consists of determining a suitable number of components with corresponding FWHM and central wavelength values.
While it is straightforward to configure PAHFIT to fit the FWHM and central wavelength between given upper and lower bounds, we aim first and foremost for a set of fixed FWHM and wavelengths that are a good enough compromise to fit all five Orion Bar templates, to facilitate a comparison of band strengths between spectra of different targets.

We made use of the atomic PDR and DF3 templates, as this pair shows the most diverse emission in the Orion Bar, and we broke down these spectra into smaller wavelength ranges that are more straightforward to work with.
The selected wavelength segments for tuning each complex are 3.1-3.6~\mum, 5.1-5.8~\mum, 5.8-6.7~\mum, 7.9-9.0~\mum, 10.5-15.0~\mum, and 15.7-18.0~\mum, for which detailed comments and visualizations are provided in Appendix~\ref{app:tuning}.
In this section, we provide an overview of general strategies that we applied.
As this is a multi-component model where changes in the component parameters may influence other components in the fit, the fine-tuning is an iterative process.

For each wavelength segment, we first removed the gas lines and narrow artifacts, then subtracted the dust continuum emission, and then (approximately) subtracted contributions by the Drude wings of dust features located outside the wavelength range.
In most cases, the density of emission lines is low enough to apply an automatic line removal that masks out any data points within a distance of two FWHM from the center of each line.
Additional lines and artifacts were removed by manually identifying and masking the problematic wavelength intervals.
For the 3.3-3.6~\mum range specifically, a manual removal was applied to all emission lines to minimize the loss of data points, as the line density is higher here.

The continuum subtraction for each wavelength segment starts by considering a preliminary PAHFIT fit using the classic science pack, with the dust continuum model replaced by our alternate model (Eq.~\ref{eq:altmbb}).
The total continuum resulting from this all-wavelength fit was then subtracted from the data.
The same classic fit was then used to calculate the sum of all dust features, excluding those located in the wavelength range of of interest. 
By subtracting this sum, we removed the contributions from the wings of wide features outside the range of interest.
In cases where the classic pack is not accurate enough to fully apply the subtractions above (e.g., Appendix~\ref{sec:filler}), the data were further rectified by subtracting a linear function that intersects the first and last point of the spectrum.

With the above cleaning steps, the remaining data should represent pure AIB spectra, and these were then normalized to their local maximum, setting the peaks of the brightest profiles to unity.
Identifying the number of features is straightforward in most cases, thanks to the spectral resolution and very high S/N of the Orion Bar spectra, and we added one feature for each visible local maximum in the spectrum.
Most of the initial identifications to determine the number of features were based on the detailed figures and list of wavelengths of \citet{Chown2024}.

After identifying the number of Drude profiles, initial wavelengths and FWHM were assigned based on the values from the classic science pack, or visually estimated from the local maxima.
An initial optimization run allowed the FWHM to vary between 0.5 and 1.5 times their initial values, while the central wavelengths are kept close to the observed local maxima, and the Drude amplitudes vary between 0 and 1.
We then iterated this approach manually, adding a minimum number of subcomponents to reflect all local maxima and inflections and adjusting the initial values and bounds until the absolute residuals were smaller than 0.05.
We note that the observational uncertainties on the spectra are generally very small for the template spectra, which may lead to overfitting.
For the tuning based on these individually cleaned wavelength segments, we set the uncertainty to $\sigma_y = 0.01 y + 0.01$, where $y$ is the normalized intensity of the cleaned spectral segment.
The performance of the tuning is validated in Sect.~\ref{sec:tuningvalidation}, where residuals are shown.

As the above approach was applied to both atomic PDR and DF3 template spectra, two independent tuning solutions were obtained, and together these indicate a plausible range for each parameter.
To construct a PDR science pack in which the FWHM and central wavelengths are constant for most dust features, we need to select suitable values in between the atomic PDR and DF3 tuning results.
The selections for the parameters were made for each feature separately and, in most cases, the mean values were used.
In cases where a feature is weak or ambiguous in one of the template spectra, we used the FWHM and wavelength results of the template where the feature was most pronounced. 
For a detailed account of the tuning process and the decisions we made for each group of emission features, supporting figures, and an overview table, we refer to Appendix~\ref{app:tuning}.
The number of AIB components in the PDR pack (66) is about twice that of the classic pack (26).

\section{Results}
\label{sec:results}

\subsection{Application to the Orion Bar templates}
\label{sec:tuningvalidation}

\begin{figure*}[htb]
    \centering
    \includegraphics[scale=0.9]{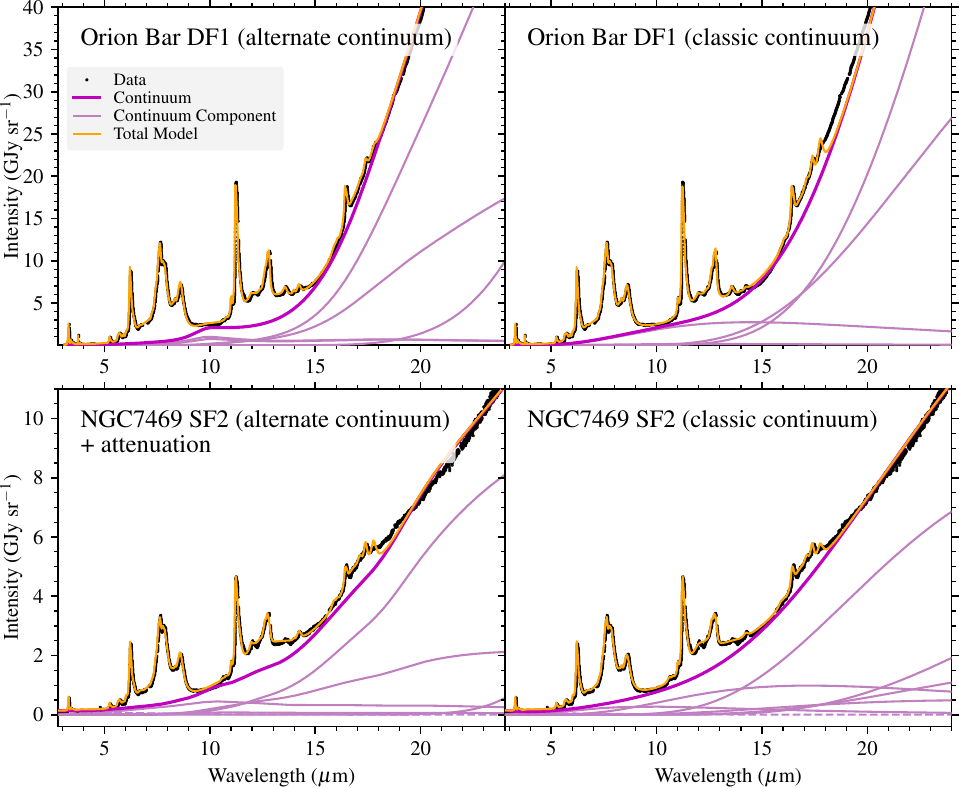}
    \caption{Comparison of the two continuum models when applied to both the Orion Bar and NGC7469.
    The PDR pack was applied to the full wavelength range, with different continuum configurations as labeled.
    For the Orion Bar DF1 template spectrum, the main differences occur near 7.7, 13, and 18 \mum.
    Regardless of the continuum used, the fitted attenuation is zero.
    For the NGC7469 SF2 spectrum, using the classic continuum results in zero attenuation, while the alternate continuum results in an attenuation of $\tau_{9.7} = 1.5$, which flattens out the silicate emission features.
    A zoomed in view and residuals of the DF1 (alternate continuum) and SF2 (classic continuum) fits are shown in Figs.~\ref{fig:residualdf1} and~\ref{fig:residualsf}.}
\label{fig:continuum}   
\end{figure*}

\begin{figure*}[tp]
\centering
    \includegraphics[scale=0.9]{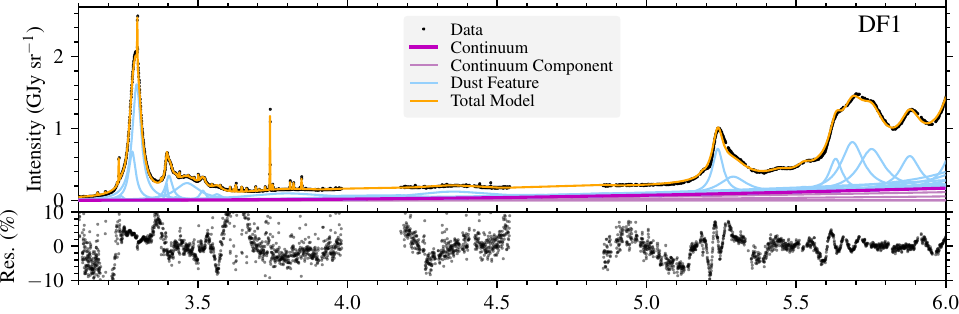}
    \includegraphics[scale=0.9]{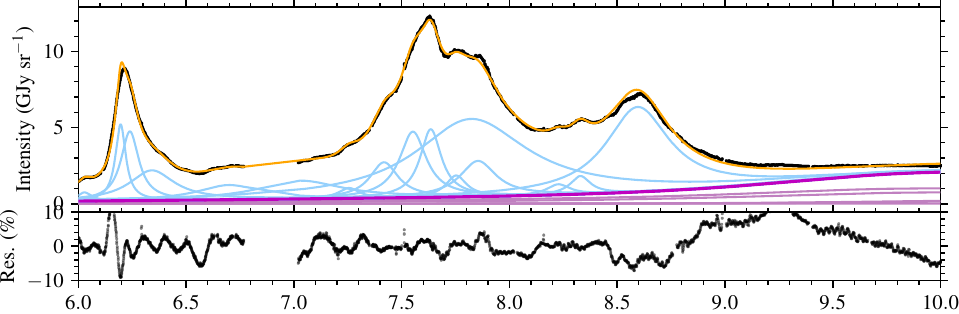}
    \includegraphics[scale=0.9]{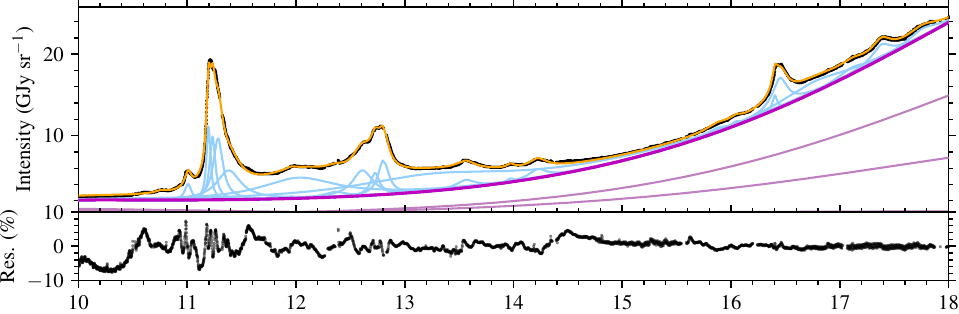}
    \includegraphics[scale=0.9]{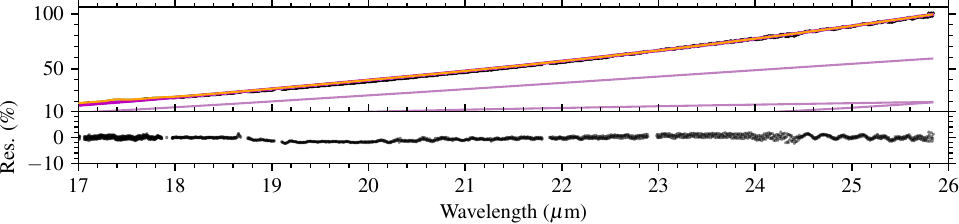}
    \caption{Illustration of fit with PAHFIT and the PDR pack applied to the Orion Bar DF1 template spectrum over the 3.1-26~\mum wavelength range using the alternate continuum model.
    The gaps in the data are wavelength sections that were omitted (Sect.~\ref{sec:cleandata}).
    The lower part of each panel shows the fractional residuals.
    The fits to narrow lines in the 3.1-3.8 \mum range are included in the total model curve but not shown individually.
    The uncertainties on the data are not shown as they are very small, and the residuals are outside the $1 \sigma$ range for most of the data points.
     }
    \label{fig:residualdf1}
\end{figure*}

Here, we provide an overview of the fitting outcome of the PDR pack and both the standard and alternate continuum models (Sect.~\ref{sec:pahfitcontinuummodel}) when applied to the full 3.1-26~\mum spectra.
For this purpose, PAHFIT was applied to all five Orion Bar templates, and we show the results for DF1.

First, we demonstrate the behavior of the continuum models in Fig.~\ref{fig:continuum} (top panels).
We note that the continuum near 3~\mum is likely underestimated since no component representing scattering or nebular continuum emission was added (see also Sect.~\ref{sec:caveatcontinuum}).
While 13 temperatures were specified in Sect.~\ref{sec:pahfitcontinuummodel}, only a few of them have major contributions in each fit result.
For the DF1 fit shown Fig.~\ref{fig:continuum} the main contributors are the 60, 90, 280, and 360~K components for the alternate continuum.
For the fit with the standard continuum, we refined the original set of temperatures to the following: 50, 58, 65, 78, 90, 113, 135, 168, 200, 250, 300, 400~K.
The main contributors in this case are the 65, 78, 200, and 250~K components.
Despite this refinement, the standard continuum results in a large mismatch near 20~\mum (Fig.~\ref{fig:continuum}, top-right panel), as expected from our demonstration in Sect.~\ref{sec:pahfitcontinuummodel}. 
For both continuum models, the resulting attenuation is zero for all templates despite previous observational results that indicate significant attenuation of the H$_2$ lines \citep{VanDePutte2024}.
More discussion about the impacts of the continuum model and the issues with the attenuation is given in Sect.~\ref{sec:attenuation}.
For the rest of the discussion of the Orion Bar templates, we used the fit results with the alternate continuum.

To show the contents of the PDR pack and to evaluate the fitting performance, we examined the residuals over all wavelengths in Fig.~\ref{fig:residualdf1}.
The residuals (relative deviations) are well within 5\% for most of the wavelength range, with occasional excursions to the order of 10\%.
Similar performance was achieved for the four other PDRs4All template spectra, demonstrating the flexibility of the science pack to adequately fit spectra from the atomic PDR to molecular PDR regime.

The fits used for the tuning, with fully flexible FWHM and central wavelengths, can match the processed data very well over smaller wavelength ranges (Appendix~\ref{app:tuning}).
In contrast, the fit using the PDR pack over the full wavelength range uses mostly fixed FWHM and central wavelengths and results in a few problem areas.
The main issues appear with the weakest features, as they are more sensitive to the continuum and the wings of nearby stronger features.
While a tuning is provided for those weak features for completeness, the rest of this work considers the main bands.

We conclude that the new science pack provides a detailed decomposition suitable for the PDRs4All data and approximates the spectra to within a few percent. 
This should be sufficient to probe the strengths and profile variations of most PDRs observed with JWST in the Galaxy.
Considering the compromises made for the PDR pack, there may be particular science cases where the fit results are not precise enough, such as those requiring measurements of particularly weak features or more subtle AIB profile variations.
In such situations, the PDR pack functions as a starting point for tailored science packs with modified or flexible FWHM and wavelengths for specific AIBs.
For spectra averaged over large spatial regions in external galaxies, more feature uniformity and kinematic broadening is expected.
Additional science packs for such environments are forthcoming.

\subsection{Application to the NGC7469 star forming ring}
\label{sec:applyngc7469}

\begin{figure*}[tp]
    \centering
    \includegraphics[scale=0.9]{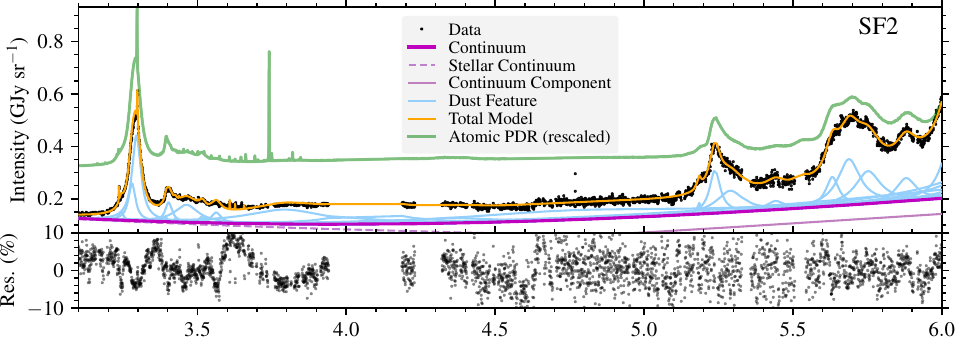}
    \includegraphics[scale=0.9]{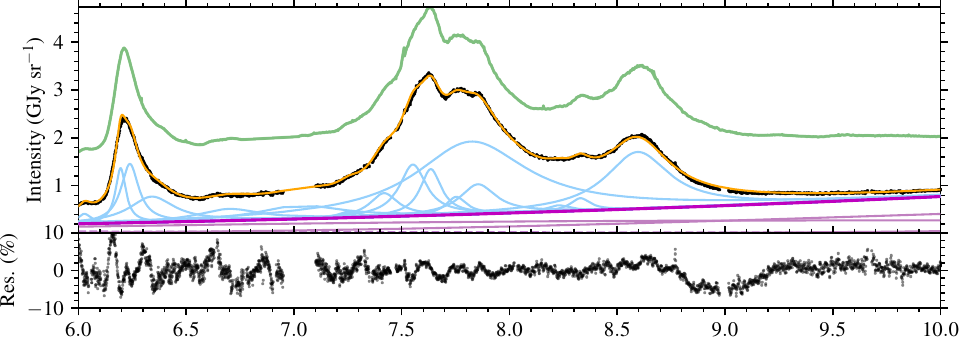}
    \includegraphics[scale=0.9]{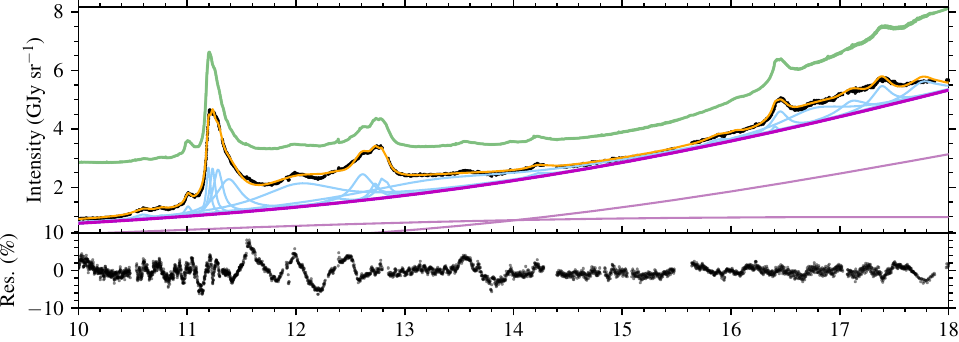}
    \includegraphics[scale=0.9]{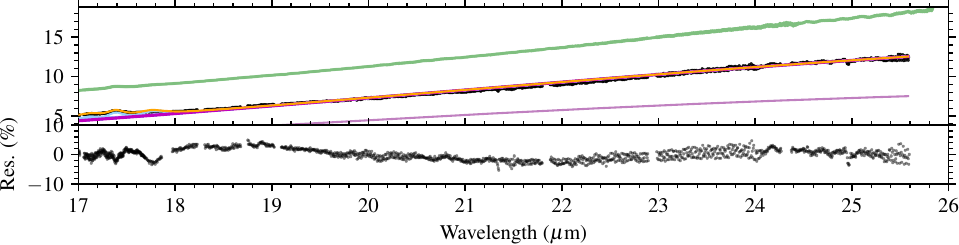}
    \caption{Application of PAHFIT and the PDR pack to the NGC7469 SF2 spectrum, using the classic continuum.
    For comparison, a rescaled and shifted version of the Orion Bar atomic PDR template spectrum is shown in green.}
    \label{fig:residualsf}
\end{figure*}

We applied PAHFIT and the PDR pack to the NGC7469 spectra with both the standard and the alternate continuum models, and found that both fits resulted in comparable residuals (see Fig.~\ref{fig:continuum}).
We adjusted the temperature set for the standard MBB components to the following: 35, 65, 78, 90, 113, 168, 200, 300, and 400 K.
To fit galaxies, a stellar continuum component is needed as well, to account for the stars that are in the same beam as the dust.
This is approximated with a 5000~K blackbody (regular BB, not a MBB), analogous to IDL PAHFIT or the classic pack for Python PAHFIT.

The most important difference between the fits with the two continua is the resulting attenuation, as using the standard continuum results in $\tau_{9.7} \approx 0$, while using the alternate continuum results in $\tau_{9.7}$ values of 0.9-1.8.
This is unsurprising, as the fitter strengthens the attenuation to compensate for the additional silicate emission included in the alternate dust continuum curves.
This represents a potential fitting instability, and a degeneracy between the choice of the dust continuum and attenuation curve shapes (see also Sect.~\ref{sec:caveatcontinuum}).
This behavior contrasts with the fit results for the Orion Bar spectra, where $\tau_{9.7}$ was zero when using either continuum model.

We note that the alternate continuum was specifically developed to address the long-wavelength behavior of the Orion Bar spectra, and is expected to work better than the default continuum in star forming regions with a steep continuum rise, whereas
the default model is preferred for continuum emission that increases more gently at longer wavelengths.
With this in consideration, we adopted the standard continuum for the rest of this demonstration and the analysis of NGC7469 in this work.

An overview of the fit results and residuals for the NGC7469 SF2 spectrum is given in Fig.~\ref{fig:residualsf}.
The Orion Bar 
atomic PDR spectrum is also shown to emphasize its similarity.
Inspecting the residuals reveals at most a 10\% difference across most of the wavelength range, a performance similar to that for the Orion Bar spectra.
This indicates that the PDR pack, which is based on a Galactic PDR (the Orion Bar), is a good starting point for fitting external galaxies.
This was expected considering the similarity between the NGC7469 and Orion Bar spectra (Fig.~\ref{fig:residualsf}). 
In other words, to fit the spectra of external galaxies, finding the right continuum and attenuation prescription will take priority over further fine-tuning of individual AIB components.

\section{Discussion}
\label{sec:discussion}

\subsection{Diagnostics to quantify AIB profile variations}

To date, PAHFIT has been primarily used (and has been proven to be powerful) to obtain strength ratios between the different AIB bands. 
This topic is not discussed in this work, as it is better to address this with a larger set of astronomical objects.
Even with JWST, less detailed decompositions such as those using the classic pack can still be used for this purpose \citep[e.g.,][]{Maragkoudakis2018, Maragkoudakis2025}.
Here, we aim to extend the power of PAHFIT and present additional (complementary) diagnostics to probe the AIB profiles, facilitated by JWST's superb spectral resolution and the new PDR pack presented in Sect.~\ref{sec:results}. 

Other works in the PDRs4All series already discuss several AIB profile differences.
\citet{Chown2024} address the profile variations of all the main complexes, in a qualitative way, and a data-driven morphological classification of the shape of the spectrum and the corresponding spatial zones is provided by \citet{Pasquini2024}.
Detailed discussions of specific regions include work on the 3.3-3.6 \mum region \citep{Peeters2024, Schroetter2024}, and the 10-15 \mum region \citep{Khan2025} 
while Schefter et al. (in prep.) investigate the average size of the PAH population and its relation to the AIB profiles.
In these previous works, the widths of the 3.3, 3.4, 5.2, 6.2, 7.7, 11.2, and 12.7 \mum features are found to broaden across the template regions in the following order: atomic PDR, DF1, HII, DF2, DF3.
In a similar order, the prominence of certain subcomponents contributing to the 5.7, 7.7, and 11.2 \mum profiles becomes weaker in the surface layers of the PDR. 
These changes are attributed to the removal of certain small and relatively unstable carriers through photolysis by a harsher radiation field
(see also \citealt{Peeters2024, Schroetter2024, Pasquini2024, Khan2025}; Schefter et al. in prep.).

\begin{figure*}[htbp]
    \centering
    \includegraphics[scale=0.9]{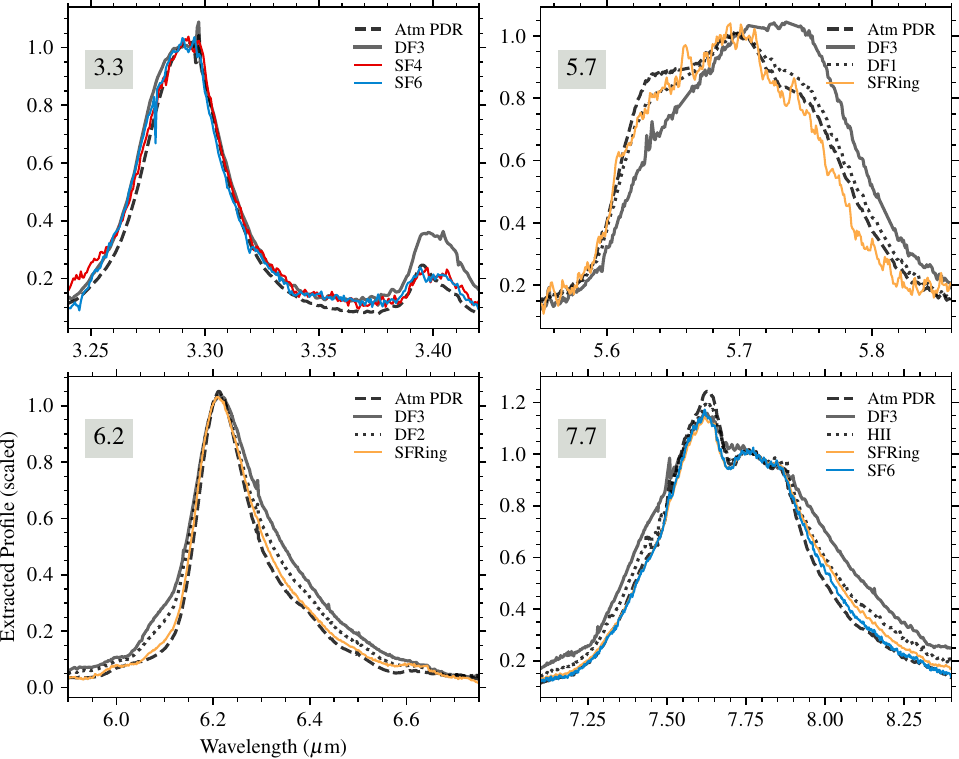}
    \caption{Comparison of extracted AIB profiles.
    The PAHFIT fits were used to compute and subtract all components outside of the feature of interest (continuum, other AIBs).
    Each profile ($F_\nu$ units, MJy sr$^{-1}$) is normalized to the average intensity in a small wavelength window, which is near the peak for 3.3 and 6.2~\mum, at 5.7~\mum, or at 7.75~\mum.}
    \label{fig:extraction}
\end{figure*}

We aim to quantify the profile variations of the 3.3, 3.4, 5.7, 6.2, and 7.7~\mum bands.
For each band, we first provide a brief description the typical assignment, and the observed profile changes and their possible interpretations, and then propose a diagnostic based on PAHFIT to quantify those changes.
To support the discussion of the following sections, we also provide a visual comparison of the profiles extracted from the data using PAHFIT (Fig.~\ref{fig:extraction}).
This extraction was performed by computing the sum of all PAHFIT model components that do not contribute to the profile of interest, and subtracting this sum from the data.

\subsubsection{The two-component 3.3 and 3.4~\mum profiles}
\label{sec:diag33}

The 3.3~\mum band originates from aromatic C-H stretching modes, and shows relatively small profile variations even under widely different physical conditions \citep{Tokunaga1991, vanDiedenhoven2004}.
While the broadening of a PAH emission band can typically be explained in terms of the PAH excitation temperature and anharmonicity in the vibrational transitions \citep{Joblin1995,Pech2002,Mackie2022}, the changes seen in experiments and theoretical calculations of small PAHs are much larger than what is observed in space.
The existence of two components in the observations, with a connection to different edge structures, was shown by \citet{Song2003, Song2007} and \citet{Candian2012}, though this is for the class B AIBs observed in the Red Rectangle, while all 3.3~\mum AIBs in the Orion Bar spectra are of class A.
The JWST observations of the Orion Bar clearly show the presence of (small) profile variations within class A. Indeed, there is a subtle change in the width of the 3.3~\mum band, along with a small shift in the blue direction \citep{Peeters2024, Chown2024}. Further investigation is warranted to explain the discrepancy between the observations and the experimental and theoretical spectra.

The 3.4~\mum feature is typically assigned to the presence of carbonaceous molecules with aliphatic bonds. Suggestions for such carriers include molecules with alkyl side groups \citep[e.g.,][]{Geballe1989, Joblin1996, Yang2013, Yang2017, Maltseva2018, Buragohain2020} or superhydrogenated PAHs \citep[e.g.,][]{Yang2020}.
Profile variations of the 3.4~\mum feature were discovered via ground-based observations of the Orion Bar \citep{Sloan1999}.
\citet{Peeters2024} and \citet{Chown2024} show that the 3.4~\mum profile has a sharp peak on the blue side in the atomic PDR, while in DF3 the profile has a nearly flat top.
The decomposition and maps by \citet[][Fig.~17]{Peeters2024} reveal that a component on the red side is strongly enhanced in DF3 compared relative to the 3.3~\mum map, suggesting that the 3.4/3.3 ratio variations are driven by this subcomponent.

In the PDR pack, both the 3.3 and 3.4~\mum features are modeled with two components (3.280 and 3.296 \mum for 3.3~\mum; 3.395 and 3.403 \mum for 3.4~\mum; see Appendix Fig.~\ref{fig:tuning3336} and Table~\ref{tab:tuning}).
We propose to trace the broadening toward the blue wing of the 3.3~\mum profile using the fractional contribution of the blue component to the total profile, which we call 3.3A.
Analogously, we trace the widening and flattening of the 3.4~\mum peak using the fractional contribution its red component, which we call 3.4B.
Both quantities are shown in the left panel of Fig.~\ref{fig:shapesmerged}.

\subsubsection{The three-component 5.7~\mum profile}

In the Orion Bar templates, the 5.7~\mum profile has three local maxima, of which the blue component is more prominent for the atomic PDR, while the red component is enhanced for DF3 \citep{Chown2024}.
The 5.7~\mum feature and others in the 5-6~\mum range likely arise from overtones or combination bands of the C-H out-of-plane bending modes that produce the 11.2~\mum complex \citep{Allamandola1989, Boersma2009, Mackie2015, Chown2024}.
For this reason, it is suggested that they trace the neutral PAH population.
Simulated spectra of small PAHs indicate that different classes of edge structures, as defined by the type of hydrogen grouping (solo, duo, trio, quartet) produce bands clustered around different wavelengths in the 5.2-5.8~\mum region \citep{Boersma2009}.
A shift in the prominence of the three 5.7~\mum subcomponents may therefore trace a change in the structure of the PAHs, in terms of these hydrogen adjacency classes. 

The 5.7~\mum complex is described by three components in the PDR pack, at 5.63, 5.69, and 5.75 \mum.
We use the fractional contributions by the blue, middle, and red components relative to their sum, and refer to them as 5.7A, 5.7B, and 5.7C respectively.
We show all three values in the middle panel of Fig.~\ref{fig:shapesmerged}, where the horizontal and vertical axes show 5.7A and 5.7C, while the 5.7B value is represented by a grid of dashed lines (as $A + B + C = 1$).

\subsubsection{Width of the 6.2~\mum profile}

The 6.2~\mum feature is associated with pure aromatic C-C stretching modes, and is stronger for cationic PAH species \citep{Allamandola1999, Peeters2002}.
\citet{Chown2024} found that its profile broadens both on the blue side and in its extended red wing.
This broadening is connected to the excitation temperature of the carriers via the effects of anharmonicity \citep{Mackie2022} and may therefore trace small PAHs, as these can reach higher excitation temperatures compared to larger PAHs upon absorption of a given photon energy.

There are three components that contribute to the 6.2~\mum feature in the PDR pack (Figs.~\ref{fig:residualsf},~\ref{fig:tuning62}). 
The first two are near the peak (6.196, 6.239~\mum), and these are the only components in the science pack for which a flexible FWHM was used (Table~\ref{tab:sums}).
The third component (6.341~\mum) represents the extension of the red wing, and has a fixed FWHM.
To measure the broadening of this profile, we used the PAHFIT results to sum the Drude profiles of all three components, and numerically computed the FWHM based on the resulting curve.
We explicitly list the components used in this sum in Table~\ref{tab:sums}.

\subsubsection{Broad and narrow components of the 7.7~\mum profile}

The multiple peaks of the 7.7 \mum band were already revealed by ISO observations \citep[e.g.,][]{Peeters2002}, and these are assigned to modes with a mixed character of C-C stretching or C-H in-plane bending vibrations.
The narrower peaks (near 7.65, 7.75, 7.85~\mum) and a shoulder-like feature in the blue wing (near 7.45~\mum), are most prominent for the atomic PDR template and least prominent for the DF3 template, while the complex as a whole is narrowest in the atomic PDR, and broadest for DF3 \citep{Chown2024}.
At lower spectral resolutions, the 7.7 \mum profile was previously interpreted as having two main components named ``7.6'' and ``7.8'' \citep[e.g.,][]{Bregman1989, Cohen1989, Molster1996, Roelfsema1996, Moutou1999, Moutou1999a, Peeters1999, Bregman2005, Berne2007}.
A broad component near this wavelength may be related to PAH clusters or very small grains (VSGs) \citep{Berne2007, Pilleri2012, Peeters2017, Stock2017, Khan2025}.

\begin{figure*}[htb]
    \centering
    \includegraphics[width=0.320\linewidth]{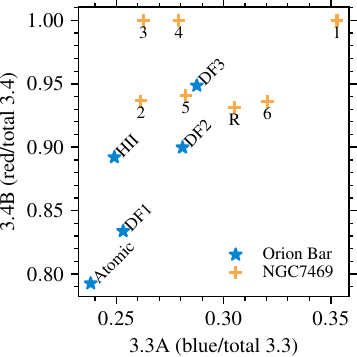}
    ~ %
    \includegraphics[width=0.320\linewidth]{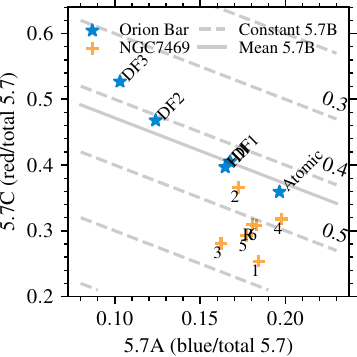}
    ~ %
    \includegraphics[width=0.320\linewidth]{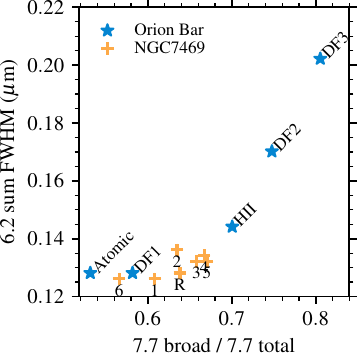}
    \caption{Profile diagnostics derived from the PAHFIT fits of the Orion Bar templates and the NGC7469 SF spectra.
    The numbers 1-6 refer to SF1 through SF6, and `R' refers to  SFRing.
    Left: Relative contribution of the blue component of the 3.3~\mum profile (3.3A), and the red wing of the 3.4~\mum profile (3.4B).
    Center: Relative contributions of the blue (5.7A), center (5.7B), and red (5.7C) components of the three-component 5.7 \mum profile.
    The dashed lines represent a grid of constant 5.7B, and the solid line indicates the mean (0.42) over the Orion Bar template spectra.
    Right: FWHM of the summed 6.2~\mum complex versus relative contribution of the broad component to the 7.7~\mum complex.}
    \label{fig:shapesmerged}
\end{figure*}

To quantify the prominence (or lack thereof) of the narrow subcomponents of the 7.7~\mum feature, we computed the fractional contribution of the broad feature near 7.9~\mum (``7.7 broad''), relative to the total power of all components between between 7.4 and 7.9~\mum (six narrow ones + ``7.7 broad'' = ``7.7 total'').
The broad Drude component included in the PDR pack may play a similar role as the ``7.8'' component does in the aforementioned studies.
An explicit listing of the components that contribute to ``7.7 total'' is given in the appendix (Table.~\ref{tab:sums}).
This ``7.7 broad'' diagnostic is shown in the right panel of Fig.~\ref{fig:shapesmerged}, where it is compared to the ``6.2 FWHM'' diagnostic proposed in the previous section.

\subsection{Using profile diagnostics to trace the photochemical evolution of PAHs}

\subsubsection{Diagnostics applied to Orion Bar}
\label{sec:diagnosticsorion}

Here, we discuss the results for the profile diagnostics described in the previous section, when applied to the Orion Bar.
In the left panel of Fig.~\ref{fig:shapesmerged}, we observe a progression from low to high 3.3A values in the order of atomic PDR, HII $\approx$ DF1, DF2, DF3, showing that this diagnostic correctly traces the sequence in which the 3.3~\mum profile broadens.
Analogously, the 3.4B diagnostic represents the order in which the red component of the 3.4~\mum becomes more prominent \citep{Chown2024}. 
The difference between the two extreme cases, atomic PDR and DF3, is shown in the upper left panel of Fig.~\ref{fig:extraction}.
The 3.3A and 3.4B values for the HII template do not exactly match the above sequence, because of biases related to the nebular continuum of the ionized gas near 3.0~\mum (see Sect.~\ref{sec:caveatcontinuum}). 
In addition, the fit residuals of the numerous stronger \ion{H}{i} recombination lines in this region may play a role as well (Sect.~\ref{sec:caveatlines}). 
We further note that the AIB emission in the HII template arises from the face-on background PDR whereas the other 4 templates probe individual key zones within an edge-on PDR.
The variation of 3.4B is large, spanning a range of 80-95\% in its contribution to the total power of the profile, while the contribution by 3.3A stays within a narrower range.

For the 5.7~\mum diagnostics (Fig.~\ref{fig:shapesmerged}, middle panel), the 5.7A and 5.7C fractions vary the most (by nearly a factor of 2), while 5.7B stays relatively close to the mean (0.42, solid gray line). 
This shows that the profile changes are mostly described by a progressive exchange in surface brightness from the blue to the red component, in the sequence: atomic PDR, DF1 $\approx$ HII, DF2, DF3. The change in the 5.7~\mum profile is illustrated in Fig.~\ref{fig:extraction} for the extremes of this sequence.

The ``7.7 broad / 7.7 total'' and ``6.2~FWHM'' diagnostics are shown in the right panel of Fig.~\ref{fig:shapesmerged}.
The same sequence is retrieved for both quantities, increasing in the following order: atomic PDR, DF1, HII, DF2, DF3.
The extracted atomic PDR and DF3 profiles for these quantities are shown in Fig.~\ref{fig:extraction} to clarify the extremes of this sequence.

In all cases, our diagnostics preserve the ordering of the templates identified visually by \citet{Chown2024}.
Previous classification work of the spectral variability in the Orion Bar provides a consistency check.
\citet{Pasquini2024} determine key spatial regions and their associated 3.2-3.6, 5.95-6.6, and 7.25-8.95 profiles using k-means clustering, and find similarly ordered progressions, including the similarity of HII region to the DF1 or DF2 regions.
\citet{Schroetter2024} use a principal component analysis to derive an ``irradiated'' and a ``shielded'' template for the 3.2-3.6~\mum complex. 
In the Orion Bar, this ``irradiated'' template dominates atomic PDR region, while the HII template region has an excess contribution by the ``shielded'' template, consistent with its position in between the DF1 and DF2 templates in our diagnostic diagrams.
This shows that with the help of the PAHFIT decomposition of JWST data, the profile differences of AIBs observed in PDRs can be quantified, and compared to the physical cases represented by the Orion Bar templates.

When not due to blending of multiple bands, the width of AIB profiles is set by the excitation temperature of the emitting PAHs and anharmonicity \citep{Joblin1995,Pech2002,Mackie2022} with the excitation temperature being set by the size of the PAH molecule and the absorbed photon energy \citep[e.g.,][]{Leger1989, Allamandola1989}. 
\citet{Chown2024} and Schefter et al. (in prep.) report the average PAH size in the Orion Bar to be largest near the surface of the PDR, in the atomic PDR, and decreasing with depth into the PDR (i.e., in the direction DF1, DF2, and then DF3). 
They conclude that the AIBs' FWHM increases in more shielded environments, because the smaller PAHs that are present in these shielded environments can reach higher excitation temperatures \citep[][Schefter et al. in prep.]{Chown2024}.
The three diagnostics, 3.3A, 3.4B, and 6.2 FWHM, presented here all quantify this decrease of the average size of the PAH population in the Orion Bar and thus quantitatively probe the chemical evolution of the PAH population due to UV photolysis. 

The 5.7A and 5.7C diagnostics also change monotonically along this sequence, and this may encode additional information about the photochemical evolution of their carriers.
The specific vibrational assignment of the 5.7~\mum AIB components is still under investigation, though clues may lie in its connection to the 10-15~\mum complex.
\citet{Khan2025} present a detailed decomposition of the 10-15 \mum bands for the Orion Bar, and maps of the 10.8, 11.0, 11.207, 12.7, 13.5, and 14.2~\mum AIBs associated with the C-H out-of-plane bending modes.
These authors conclude that the PAH edge structures, in terms of their hydrogen adjacency classes, are dominated by solo and trio C-H groupings, while duo C-H groups are more abundant deeper in the PDR.
The 5.7~\mum band is thought to consist of the overtones of some of the 10-15 \mum bands \citep{Allamandola1989, Boersma2009, Mackie2015}, and its variations may therefore trace similar structural changes.
In particular, the carriers responsible for the increase in the red component (5.7C) should prefer more shielded environments and be suppressed by stronger radiation fields, while the material associated with the blue component (5.7A) should be more resilient to the conditions near the surface of the PDR.
The PDR pack provides the opportunity to measure the subcomponents of both the 5.7 and 10-15~\mum complexes in a single decomposition, which will deepen the potential connections by revealing which subcomponents best correlate across the two complexes.

Our last diagnostic, the broad~7.7 / total~7.7 ratio, also exhibits a monotonic change over the different key zones in the PDR, being strongest in the more shielded environments. 
We hypothesize that this diagnostic traces (part of) the fractional contribution to the 7.7~\mum emission from a component consisting of PAH clusters or very small grains (VSGs) \citep{Berne2007, Pilleri2012, Peeters2017, Stock2017, Khan2025}. \citet{Berne2007} and \citet{Pilleri2012} reported that the VSG contribution is prominent in more shielded environments as VSGs photo-evaporate under the influence of UV radiation.
The observed relationship between the 6.2 FWHM and the broad~7.7 / total~7.7 diagnostic can then be understood as both diagnostics increasing in more shielded environments. 

Since each template occupies a specific region in the diagnostic diagrams along a monotonic sequence, this raises the opportunity to classify the AIB profiles observed in other objects. 
Via the similarity of an AIB profile to one of the Orion Bar templates, which span a range of photochemical evolution stages as described above, one could identify a region in the Orion Bar that is a good analog for the physical conditions giving rise to an observed AIB spectrum.
The study of 30 Dor by \citet{Zhang2024} finds that those JWST spectra are similar to the DF2 Orion Bar template in terms of the AIB profiles, while the band strength ratios show a larger range of variability, indicating that the profile diagnostics may complement the conventional diagnostics based on band ratios.

\subsubsection{Diagnostics applied to NGC7469}
\label{sec:diagnostics7469}

We applied the same diagnostics to the SF spectra of NGC7469 (Fig.~\ref{fig:shapesmerged}, orange markers).
A selection of extracted AIB profiles were added to Fig.~\ref{fig:extraction} as well, to show if the position in the diagrams is consistent with what we see in the extracted profiles.
The 3.3A diagnostic spans a range almost twice that found for the Orion Bar.
Despite this, the extracted profiles of Fig.~\ref{fig:extraction} show a minimal change in width (SF4 and SF6 shown as examples).
Instead, a slight shift in the central wavelength is seen when inspecting the wings, which can be attributed to the different line-of-sight velocities at different positions in the starburst ring.
For the 3.4A diagnostic, several of the values are zero, indicating the sensitivity of the diagnostic to the strength of the 3.4~\mum band and the signal-to-noise ratio of the data. Indeed, the 3.4~\mum profiles (Fig.~\ref{fig:extraction}) are shaped differently but they are noisier.
Extragalactic observations are subject to stronger velocity effects, and a precise redshift and velocity correction may be required to apply these metrics.
The presence of emission lines is a particular issue at 3.4~\mum (see also Sect.~\ref{sec:caveatcontinuum}).
Even with velocity corrected data and high S/N, the observations are averaged over much larger spatial scales when external galaxies are observed, which results in more consistent profiles.
This emphasizes that extremely deep and spatially resolved observations are needed to study the behavior of these specific profile variations, which is enabled by JWST.

The 5.7~\mum diagnostics are clustered in a region with 5.7A values between those of the Orion Bar atomic PDR and DF1/HII, with higher 5.7B and lower 5.7C values.
This describes the behavior seen in the upper right panel of Fig.~\ref{fig:extraction}, where the 5.7~\mum profile of the SFRing spectrum resembles that of DF1, but with a slightly lower red component (the profiles were normalized at the central peak).
This confirms that the 5.7A, 5.7B, and 5.7C metrics can relate the profiles observed in these extragalactic star forming regions to one of the Orion Bar templates, even though the NGC7469 data are somewhat noisier at 5.7~\mum.

For NGC7469, the 6.2 FWHM is similar to that of the Orion Bar atomic PDR/DF1 templates for all seven spectra, and this is also reflected in the extracted 6.2~\mum profile (Fig.~\ref{fig:extraction}, bottom-left panel).
The broad~7.7 fraction spans a range between the atomic PDR and the HII values.
The extracted profiles (Fig.~\ref{fig:extraction}, bottom-right panel) show that the red wing of the SFRing 7.7~\mum profile is very close that of the Orion Bar HII template, while the SF6 profile is somewhere in between the atomic PDR and HII profiles, matching what we see in the diagnostic diagram.

Based on our diagnostics, the NGC7469 AIB profiles are most similar to the atomic PDR template for 5.7 and 6.2~\mum, while there is some uncertainty for the 7.7~\mum band, as the points in the diagram are spread between the atomic PDR and HII cases. 
While the precise assignment to the atomic PDR/DF1/HII template may not always be clear for extragalactic spectra, we note that DF2 and DF3 are far removed from the NGC7469 SF spectra in the diagnostics diagram (except for the problematic 3.3A and 3.4B results in this case). 

This is consistent with the PAH characteristics in NGC7469 JWST spectra reported previously. 
\citet{Schroetter2024} set up a two-component model to fit the 3.2-3.6~\mum AIBs with a linear combination of their ``shielded'' and ``irradiated'' shapes derived from the Orion Bar JWST spectra. 
These authors applied this model to the NGC7469 spectra, revealing a high contribution by the ``irradiated'' component, similar to their results for the Orion Bar atomic PDR template.
\citet{Lai2022, Lai2023} apply the CAFE tool \citep{DiazSantos2025}, and find that the 6.2/7.7 and 11.3/7.7 ratios change by around 30\% in the starburst ring, which indicates small changes in the size and ionization state. 
The SF1-SF6 regions in NGC7469 are also investigated by \citet{Rigopoulou2024}, where they compared these ratios to grids of theoretical PAH models to conclude that the SF regions in NGC7469 are dominated by charged PAHs.
The spectra of NGC7469, as well as other external galaxies, are expected to contain contributions from a variety of PDR conditions, considering the much larger spatial scales within a resolution element.
The previous works cited above, and the diagnostics in Fig.~\ref{fig:shapesmerged} along with the extracted profiles of Fig.~\ref{fig:extraction}, reveal that the AIB profiles observed in NGC7469 SF spectra are clustered together and are close to those of the atomic PDR or DF1 templates for the 5.7~\mum and 6.2~\mum features, and somewhere between the DF1 and HII cases for the 7.7~\mum feature.
Given that the diagnostics and profiles of the DF2 and DF3 templates deviate quite far from the atomic PDR case, the AIBs present in the DF2 and DF3 templates do not contribute strongly to the spectra of NGC7469.
As the AIB emission from a PDR is dominated by the emission from the atomic zone in the PDR \citep[e.g.,][]{Habart2024, Peeters2024}, AIB emission from external galaxies predominantly arises from the atomic HI PDR zone instead of the molecular PDR zone.
Hence, extragalactic star-forming regions at solar metallicity subjected to similar UV radiation fields as the Orion Bar will likely exhibit similar PAH characteristics as the atomic PDR in the Orion Bar. 

The high similarity of the AIB profiles in the NGC7469 and Orion Bar spectra could be interpreted as a manifestation of the ``GrandPAH'' scenario, where the UV photoprocessing under PDR conditions leads to a PAH population consisting of a relatively small number of robust species \citep{Andrews2015}.
The AIB profiles emitted by these stable GrandPAH species are assumed to be insensitive to the radiation field.
The remaining profile differences between NGC7469 and the Orion Bar could then be caused by different abundances of the underlying GrandPAH species.

\subsection{Range of applications}
\label{sec:applications}

Here, we elaborate on a few typical use cases we expect for the PDR pack.
First, we note that this work is not the first application of PAHFIT to JWST data.
For example, \citet{Maragkoudakis2025} apply PAHFIT with the classic pack to JWST data of two galaxies, compare the results to those of Spitzer data, and study the effects of the resolution and depth.
The use of JWST data results in a 5-10\% difference on the neutral and anion PAH fractions derived from the extracted AIBs.
The increased detail in the PDR pack will result in a more precise recovery of the AIB intensities, compared to the classic pack.
We note that IDL PAHFIT provides standardized sums to compute the total power of individual bands, to ensure these are comparable between different analyses.
Since the PDR pack is a complete overhaul and many bands have been split up into several subcomponents, we define similar sums in Table~\ref{tab:sums}.

We expect the PDR pack and its derivatives to be applicable to nearly all PDR spectra and other AIB spectra of class A observed with JWST, considering that it works well for all five Orion Bar templates, which span a relatively large range of variations within this class.
The contents of the PDR pack are focused on the AIBs, and can be combined with various setups for the stellar continuum, the dust continuum, and the attenuation model, and the last section (Sect.~\ref{sec:caveats}) elaborates on known problems and potential solutions regarding these aspects.
By addressing these aspects, the applicability of the PDR pack can be extended to low-obscuration star forming galaxies as well, considering our demonstration for NGC7469.

Now that the PDR pack is in place to make use of the spectral resolution and depth of JWST observations, a pixel-per-pixel application of PAHFIT is the next logical step to take full advantage of the spatial resolution.
A utility called PAHFITcube\footnote{\url{https://github.com/drvdputt/pahfitcube}} is under development to enable this.

\subsection{Limitations}
\label{sec:caveats}

\subsubsection{Dust attenuation}
\label{sec:attenuation}

\begin{figure*}
    \centering
\includegraphics[scale=0.9]{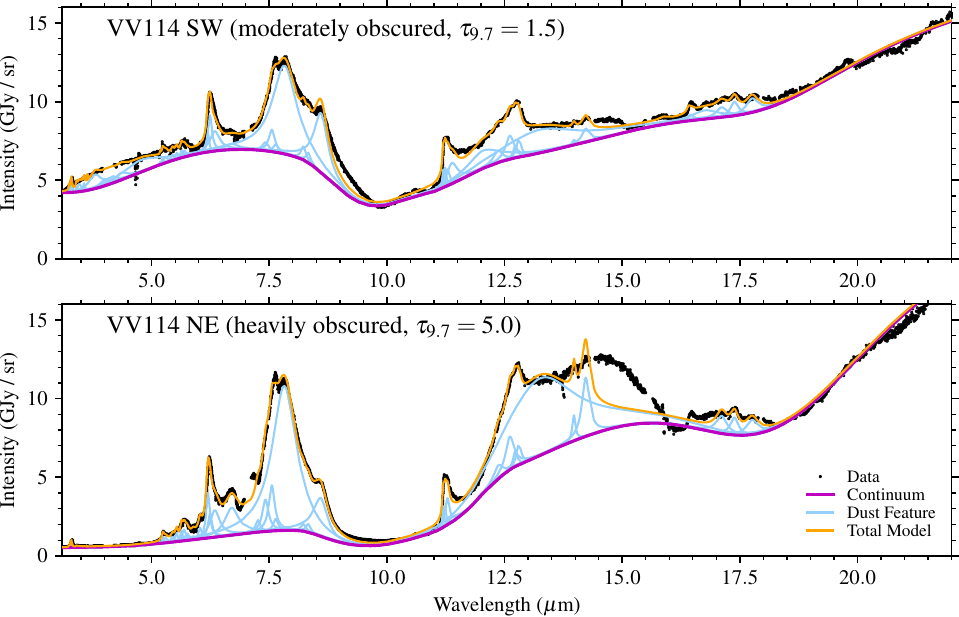}
    \caption{Application of the PDR pack to the obscured AGN of VV114.
    The corresponding apertures are analogous to \citet{Donnan2024}.
    The displayed $\tau_{9.7}$ are those resulting from PAHFIT, while \citet{Donnan2024} report total optical depths ($\tau_{9.8}$) of the order 1 and 5 for VV114 SW and NE respectively.
    Due to the limitations at higher dust attenuation, the 7.7 and 13.3~\mum broad components are likely overestimated for VV114 NE, while the mismatch near 15~\mum becomes larger.}
    \label{fig:demo}
\end{figure*}

In this section, we elaborate on the role of the attenuation curve in PAHFIT and caveats associated with it. 
The attenuation is a function of wavelength, which is applied after summing the AIB components in the model, so it differentially affects the fitted power of each AIB.
Therefore, if two fits result in different values of $\tau_{9.7}$ (e.g., when comparing different continuum models, Fig.~\ref{fig:continuum}), the derived AIB ratios will also be different.
An extinction correction using PAHFIT or other methods \citep[e.g.,][]{Stock2013} is therefore essential for obtaining comparable AIB ratios for different spectra.
In principle, different attenuation curves may be necessary depending on the object of interest, though in most cases, the same classic PAHFIT attenuation curve is used.
The effect of different attenuation geometries, meaning the relative spatial locations of the AIB emitting layers and the and extinguishing media, are explored by \citep{Lai2020}, as different assumptions are needed depending on the type of galaxy.
The MIR attenuation that affects PAH emission was empirically determined by \citet{Lai2024}, and their results favor attenuation curves that are relatively flat from 3-8~\mum, while the optical depth of the silicate features is of intermediate strength at 9.7~\mum, and relatively high at 18~\mum.
In comparison \citep[Fig.~5 of][]{Lai2024}, the classic PAHFIT attenuation curve of \citet{Smith2007} exhibits a much stronger 9.7~\mum feature relative to that at 18~\mum, and a steeper slope from 3-8~\mum.

Even with a constant shape of the attenuation curve, some degeneracy between the emission power of certain components and the silicate absorption ($\tau_{9.7}$) is common \citep[e.g.,][]{Peeters2017}.
Since our alternate continuum components contain silicate emission (see also caveats below, Sect.~\ref{sec:pahfitcontinuummodel}), and the attenuation curve contains silicate absorption, a stronger degeneracy of this kind may be present when the alternate continuum is used.
In Fig.~\ref{fig:continuum}, we applied the PDR pack with both the standard and alternate MBB continuum components.
The aforementioned degeneracy manifests in the fit results of the NGC7469 spectra with the alternate continuum, where the resulting $\tau_{9.7}$ span values ranging from 0.9 to 1.8.
For those $\tau_{9.7}$ values, the attenuation curve reduces the total intensity by about 50\% near the peak of the 10~\mum silicate feature; enough to suppress the silicate emission present in the alternate dust continuum model.
In contrast, the fits with the classic continuum result in zero attenuation, which conflicts the $\tau_{9.7}$ of around 0.4-1.0 expected from previous work \citep[e.g.,][]{Lai2022}.
So in addition to the degeneracy in the parameter space where intensity and attenuation compete, $\tau_{9.7}$ also depends on the combination of continuum and attenuation models used \citep[see also][]{Maragkoudakis2025}.
\citet{Donnan2024} developed a different approach for the attenuation, as the highly obscured galaxies in their sample are a known failure mode for the current attenuation assumptions of PAHFIT.
These authors performed a similar experiment to this work, where they compare fits with the standard MBB versus a physically motivated dust emissivity that contains silicate emission (their Sect.~5.2). 
Their results are qualitatively similar to ours: the different continua can both model the spectrum of an ordinary star forming region, but the attenuation is near zero for the standard MBB, while the inclusion of silicate emission results in non-zero attenuation.

The fits of the Orion Bar spectra result in zero attenuation regardless of the continuum model, which conflicts with previous findings based on the same data.
The foreground extinction affecting the atomic gas layer of the Orion Bar was determined to be $A(V) \sim 0.9-1.9$, the internal attenuation is $A(V) \sim 3$ to 12 depending on the region \citep{Habart2024, Peeters2024}, and \citet{VanDePutte2024} reveal a decrement of the H$_2$ $S(3)$ 9.66~\mum rotational line of a factor $\sim 0.4$, indicating that it is affected by silicate absorption.
However, those estimates are based on gas lines, and rely on their own set of assumptions about the geometry of the emitting and attenuating layers, and the extinction curve.
For objects such as the Orion Bar, it may be necessary to apply differing levels of attenuation to each component type (dust continuum, AIB).
We note that we did not apply an attenuation correction to the AIB profiles and powers for both the Orion Bar and NGC7469, as the fits used throughout the discussion all resulted in $\tau_{9.7} = 0$.

{
To illustrate the for more obscured sources, we applied PAHFIT to the two AGN regions in the merging galaxy VV114, which have different obscuration levels (Fig.~\ref{fig:demo}).
These GOALS data were reduced analogously (Sect.~\ref{sec:data}), and the extraction apertures approximate those presented by \citet{Donnan2024}.
We used the PDR pack, the classic continuum and the standard attenuation model.
For the moderately obscured VV114 SW AGN, we find that this combination produces a reasonable continuum shape with a resulting attenuation strength of $\tau_{9.7} = 1.5$, which is of the same order as the value of $\sim 0.95$ reported by \citet{Donnan2024}.
The main continuum mismatch lies just short of 15~\mum, and this behavior becomes more pronounced when the highly obscured VV114 NE AGN is considered, for which PAHFIT yields $\tau_{9.7} = 5.0$.
The mismatch near 15~\mum is now very large, making it appear as if an additional feature is needed.
However, the approach by \citet{Donnan2024} shows that fitting the continuum for this part of the spectrum is possible with the implementation of the right continuum for the AGN and suitable assumptions for the dust attenuation.
For example, the inclusion of crystalline silicate absorption features is considered necessary at high obscuration \citep{Spoon2022, Donnan2023}.
}

Separate from the dust attenuation {curve} that models silicate absorption, PAHFIT also supports individual absorption features localized around specific wavelengths.
Typically, this is used to approximate the absorption by ices, and while there is very little or no ice absorption in the Orion Bar \citep{Peeters2024}, there may be absorption in other spectra by H$_2$O (near 3 and 6~\mum), CO$_2$ (near 4.3 and 15~\mum), CO (4.7~\mum), or other ice species \citep{Boogert2015}.
A previous science pack for AKARI and Spitzer spectra of galaxies \citep{Lai2020} includes absorption features for H$_2$O, CO$_2$, and CO ice, as well as an attenuation model with a tunable geometry parameter, and these parts could be combined with the PDR science pack as a starting point.

\subsubsection{Continuum}
\label{sec:caveatcontinuum}

In the previous section, we already discussed the interaction between the continuum and attenuation models, but even without attenuation, the shape of the continuum still affects the fit results.
The demonstration of both continuum models in Fig.~\ref{fig:continuum} shows that the curvature of the total continuum is different near the 7.7-8.6 and 11.3-12.7 \mum AIB complexes, which likely leads to different results for the broad Drude profiles in those regions.

We have presented just two possible options to model the continuum, one of which was specifically created as a practical solution for fitting the Orion Bar spectra.
Variations using the same prescription as Eq.~\ref{eq:altmbb}, using a different extinction curve or dust cross section as a multiplier, may produce fits with equivalent or better residuals.
For example, for Milky Way extinction curves it is known that the strength, shape, and optical depth ratio of the 10~\mum and 20~\mum silicate features can vary between sightlines \citep{Chiar2006, Gordon2021}.
Therefore, if one were to constrain the absorption cross section or extinction curve for the dust in the Orion Bar, the relative strength of the silicate features could be adjusted. 
Fitting the silicate emission with a free parameter is possible, but this will likely lead to more degeneracies.

We note that in Eq.~\ref{eq:altmbb}, we used the same factor (extinction curve) for the MBB components of different temperatures, as a first-order approximation.
This assumption does not take into account the following physical aspects.
Firstly, both the equilibrium temperature and absorption cross section of a grain depend on its size.
Secondly, the emission below 15~\mum is dominated by stochastically heated small grains, for which a distribution of temperatures for each grain size needs to be taken into account, and the resulting continuum does not reduce to a MBB.
For these reasons, having the shape of the continuum depend on the temperature could be a useful improvement.
Third, there can be contributions by clusters, which which may result in broad plateaus, though these could be modeled as independent components where applicable (see also Sect.~\ref{sec:caveatrefinement}).

Our implementation of the alternate continuum in PAHFIT also supports a mixed use case, where the continuum type is specified for each dust continuum component individually.
This could be used to introduce a basic temperature dependency of the MBB shape where, for example, only the cool components have the silicate feature.
For the Orion Bar spectra, the alternate MBB is mainly needed for components with $T \lesssim 100$ K, as these are the main contributors to the problematic 15-22~\mum range (Fig.~\ref{fig:defaultvsspecial}). 
When we restrict the fits to the data below 15~\mum, both continuum models result in similar residuals.
We emphasize that the two continuum models presented in this work are just two defaults.
While the analysis of individual targets may benefit from changing the provided default settings, we expect the standard continuum to perform well for external galaxies and moderately illuminated nebulae, as it has done in the past.
The alternate continuum is a secondary option that can be used for targets that exhibit very steep continua.

To model the continuum near 3.0~\mum, we used the standard stellar continuum of IDL PAHFIT for NGC7469, while no continuum was used for the Orion Bar where the stellar flux is expected to be negligible.
While the choice of temperature for this model component has a minor effect at longer wavelengths (the Rayleigh-Jeans power law regime), NIRSpec also covers wavelengths below 3.0~\mum for which stellar population models may need to be implemented in the model \citep[e.g.,][]{Marshall2007}.
For the Orion Bar, we performed a fitting test that includes the  stellar component, and found that the contribution is negligible for all template spectra except HII.
In this test for the HII template spectrum, the intensity of the stellar component at 3.1-3.2~\mum is around 50\% of the extended blue wing of the 3.3~\mum profile.
This non-negligible continuum likely consists of nebular emission from the ionized gas in the HII template region, which produces a relatively flat spectrum of free-free and free-bound emission (Onaka et al. in prep.).
Other contributions may include scattered light or the effects of a ``quasi PAH continuum'' \citep{Allamandola1989, Boersma2023}.
The stellar component is not a good description for these contributions, and while the extra degree of freedom improves the fit of the 3.3-3.6~\mum complex, the tail of this blackbody also worsens the fits near 5.2 and 5.7~\mum.
Therefore, we did not use the stellar component for the Orion Bar spectra, and instead note that the 3.3A and 3.4A diagnostics (Fig.~\ref{fig:shapesmerged}) are biased for the HII template.
For Galactic star forming regions, a nebular continuum component would a useful addition to PAHFIT to solve this bias, while for external galaxies, it is likely negligible relative to the stellar emission.

\subsubsection{Gas lines}
\label{sec:caveatlines}

PAHFIT supports the fitting of gas emission lines as Gaussian components, and the instrument pack (Sect.~\ref{sec:pahfit}) was designed to optimize this by setting a fixed FWHM for each line based on the spectral resolution.
If the wavelengths are set precisely in the science pack, and the wavelength calibration of the instrument is accurate, fixed central wavelengths can be used for optimal fitting speed.
Despite the above optimizations, we opted to avoid fitting the lines where possible, as modeling the numerous (100+) lines observed in the Orion Bar \citep{Peeters2024, VanDePutte2024} often leads to  convergence problems. 
Instead, we masked out the lines in the spectra as described in Sect.~\ref{sec:tuningstrategy}. 
This made it easier to diagnose convergence issues, but also reduced the ambiguity in developing and testing the AIB configuration, as the residuals of the line fits may still introduce a minor bias.
For the 5-26~\mum range, removing the lines only results in a minor loss of data, as the spectral resolution of MIRI MRS is sufficiently high, the density of the lines is low, and the AIBs are broad.

In the 3.1-5~\mum range, the density of lines is higher, while the AIBs at 3.3-3.6~\mum are narrower. Therefore, masking out the lines results in a significant loss of data.
This is specially important for the 3.4~\mum feature, as its blue component is very narrow and there are multiple HI recombination lines or H$_2$ lines that blend with the profile.
We mitigated this issue by including the lines in the fit for the 3.2-3.6~\mum region specifically, with fixed widths and central wavelengths bounds of $\pm0.01$\%.
While most of the line flux is accounted for by the Gaussian components, we expect there may still be some residual effects on the 3.3A and 3.4A diagnostics (Sect.~\ref{sec:diag33}) for the Orion Bar HII template, since this is where the HI recombination lines are strongest and the 3.4~\mum feature is weakest.
The narrow blue component of the 3.4 \mum profile (3.4A) may exhibit some degeneracy with the H$_2$ or \ion{H}{1} lines present in its profile.

\subsubsection{Tuning}
\label{sec:caveatrefinement}

The science packs provided in this work were shown to work very well over most of the 3-26~\mum wavelength range, but we emphasize that further refinement remains possible.
Compromises were made in the tuning to obtain reasonable fits for the different profiles found across the Orion Bar templates, while preferring fixed FWHM parameters to facilitate a more direct comparison of band strengths.
Future works that focus on specific and fainter features may require adjustments to the AIB parameters in the PDR pack, or flexible FWHM parameters.
The PDR pack will be included in the open-source PAHFIT repository where it is open to community contributions.

Since our tuning method for the AIBs depends on the subtraction of the continuum as determined by PAHFIT itself (Sect.~\ref{sec:tuningstrategy}, Appendix~\ref{app:tuning}), the tuning results depend indirectly on the assumed shape of the continuum.
The sharp drop in flux on the blue side of the 10~\mum silicate emission, and the curvature underneath the 10-15~\mum complex are suspected problem areas.
We expect that these affect the tuning outcome for the 8.6~\mum feature, and the broad features at 12~\mum and 13.2~\mum.
In principle, parts of the tuning could be redone with spectra for which the continuum is less steep, which would allow the use of the classic PAHFIT continuum, to avoid some of the caveats that come with using the alternate continuum. 
A good candidate for a refinement such as this are the MIRI MRS data of the NGC7023 PDR \citep{Misselt2025}, as the continuum is much weaker and the 16-18~\mum complex is strongly pronounced \citep{Werner2004}.
Despite this, we still preferred Orion Bar spectra for their high S/N over the entire wavelength range, including the weaker 5.2 and 5.7~\mum bands.
A cross-validation of the PDR pack by applying it to different PDRs will be important, as updated packs are provided, but doing so is outside the scope of this work.

In the remaining paragraphs we address limitations related to the fact that the AIB model exclusively uses Drude profiles.
The use of Drude profiles often reduces or eliminates the need for ad-hoc (typically spline-based) continua, or the use of ``plateau'' components in addition to the smooth MBB continuum, in contrast to the case where Gaussian profiles are used \citep{Smith2007, Galliano2008}.
In reality, there are several plateau-like features associated with certain physics.
We already addressed the possible contribution of VSGs to the 7.7~\mum band \citep{Berne2007, Pilleri2012, Peeters2017, Stock2017, Khan2025}.
Other contributions are the ``quasi PAH continuum'' \citep{Allamandola1989, Boersma2023} in the 3-5~\mum range, and contributions to the 10-15~\mum and 15-19~\mum complexes \citep{Boersma2012}.
With the high resolution and sensitivity of JWST spectra, we have better constraints on the shapes of the peaks, especially since we can see the feature strengths vary over small spatial scales.
The Drude profiles that result from the high-resolution tuning are narrower, and as such their wings fill up less of the space between the peaks. 
As such, JWST spectra reveal that such plateaus may be necessary after all, even when Drude profiles are used.
This led to the inclusion of two very broad components in the PDR pack, to approximate the flux in the 10-15~\mum region (Sect.~\ref{sec:tuning1014}).
Their parameters likely depend on how the continuum subtraction was performed before the tuning process. 
A caveat with this, is that Drude profiles are not likely not a good representation for the physics that create the 10-15~\mum plateau, and the very wide wings will affect the fit results at other wavelengths (e.g., the 13.5, 14.0, and 14.2 features).
They also change in amplitude depending on the continuum model, partially compensating for the differences in the continuum baseline (Fig.~\ref{fig:continuum}).
More physically motivated plateau shapes will be needed to minimize these effects.

In this work we have not discussed any diagnostics based on the 10-15~\mum complex, which contains the most strongly asymmetric AIB (11.2~\mum).
While we provide a decomposition for it in the PDR pack, it is the complex with the most subcomponents in our tuning.
\citet{Khan2025} investigate the profile morphology of the 10-15~\mum range in detail.
Their decomposition of the 12.7~\mum band consists of six Gaussian profiles, which is of similar complexity as the four Drude components used in the PDR pack.
They also show that the feature near 12.0~\mum has a complex structure; and these details are not modeled by the PDR pack.
They decomposed the 11.2~\mum profile into two asymmetric profiles extracted empirically from the data.
In comparison, the PDR pack models the main peak with three Drude profiles, and fourth broad component is used to fit the red wing.
It is likely that the four Drude profiles used for the 11.2~\mum band could be replaced by two asymmetric profiles that approximate the empirical profiles mentioned above.
Asymmetric Drude profiles are a planned feature for PAHFIT, as many of the AIB profiles clearly reveal a steep blue side and extended red wing.
An important step forward would be to include more realistic emission profiles that properly account for the molecular physics involved in the energy cascade and specifically allow for the effects of mode interactions and anharmonicity inherent to the molecular emission process \citep{Mackie2022}.

\section{Summary}
\label{sec:conclusions}

We have presented the Python version of PAHFIT, a spectral decomposition tool originally developed in IDL for the analysis of Spitzer spectra of nearby galaxies \citep{Smith2007}. We have extended PAHFIT into the JWST era by providing new functionalities related to the fitting of the higher spectral resolution JWST spectra.
We documented the development of the first version of the ``PDR pack'' for PAHFIT, which contains settings to fit the AIBs based on the PDRs4All Orion Bar template spectra.
We evaluated the performance and applicability of the PDR pack by using it to fit all five template spectra of the Orion Bar (HII, atomic PDR, DF1, DF2, DF3) and six star forming regions in the central ring of NGC7469.
Finally, we addressed the limitations related to the current continuum and attenuation models of PAHFIT.

Our aim is to extend the power of PAHFIT and present new diagnostics derived from PAHFIT fit results that quantify the AIB profile differences and thus probe the photo-chemical evolution of the PAH population. 
We applied these new diagnostics to the Orion Bar templates and the NGC7469 star forming regions, revealing the similarities between the AIB profiles in these environments. The main results and conclusions are as follows:

\begin{enumerate}
    \item We tuned the AIB settings for the PDR pack by fitting multi-component models to the atomic PDR and DF3 Orion Bar templates.
    Based on the results, constant FWHM and central wavelength parameters were selected to obtain settings for all main AIB complexes observed in the Orion Bar spectra. 
    We listed the individual PAHFIT components contributing to each AIB complex for a consistent comparison between their band strengths. 
    \item The standard continuum model of PAHFIT is insufficient to fit the steep continuum of the Orion Bar spectra in the 15-26 \mum range, which prevents fitting the 16-18~\mum AIB complex.
    We set up an alternate continuum model that includes silicate emission, and we showed that it performs better in this regard.
    We performed our fits with both continuum models to address the limitations related to the continuum choice.
    We discussed a degeneracy between the chosen continuum model and the outcome for the dust attenuation strength when this is applied to NGC7469.
    \item With the alternate continuum and the PDR pack, PAHFIT performs well over the full 3-26~\mum range when applied to all five Orion Bar template spectra, with fit residuals within 10\%. 
    A similar performance was achieved for the NGC7469 spectra.
    \item We propose new diagnostics that quantify the profile variations of the 3.3, 3.4, 5.7, 6.2, and 7.7~\mum features. These diagnostics show a clear progression through the Orion Bar template regions, each resulting in a similar spatial sequence: atomic PDR, DF1, HII, DF2, DF3.
    Along this sequence, the blue component of the 3.3~\mum band and the red component of the 3.4~\mum band become more prominent, the contributions of the three components of the 5.7~\mum band shift more to the red component, the FWHM of the 6.2~\mum band increases, and the contribution of a wide component to the 7.7~\mum profile becomes larger.
    \item The observed sequence probes the photochemical evolution of the PAH population.
    The broadening of the 3.3, 3.4, and 6.2~\mum AIBs is ascribed to increasing excitation temperatures of smaller PAHs, which are more abundant in the shielded environments of the molecular PDR. 
    The three subcomponents of the 5.7~\mum band are carried by at least two PAH subpopulations. The one responsible for the blue component is more resilient to the conditions near the surface of the PDR, while the one responsible for the red component prefers more shielded environments.
    The wide component of the 7.7~\mum complex captures contributions by VSGs and/or PAH clusters, and it indicates their prevalence in more shielded environments.   
    \item The AIB profiles of the NGC7469 star forming region spectra closely resemble those of the Orion Bar templates, both visually and in terms of the AIB profile diagnostics.
    The best matching Orion Bar templates are the atomic PDR, DF1, and HII, while DF2 and DF3 are significantly different.
    Therefore, the NGC7469 AIB profiles likely originate from highly irradiated regions with conditions similar to the atomic layer of the Orion Bar PDR. 
    \item The PDR pack is suitable for fitting the 3-26~\mum spectra of JWST PDR observations, and this applicability likely extends to the most unobscured objects with bright AIBs of class A.
    Considering the limitations regarding the continuum and the attenuation, 
    we expect more experiments with alternative dust continua, attenuation curves, or more physically motivated AIB or plateau profiles in the near future.
\end{enumerate}

\section{Data availability}

Tables containing the updated PDRs4All template spectra (Sect.~\ref{sec:data}), the extracted NGC7469 spectra, and a machine readable version of Table~\ref{tab:tuning} are only available in electronic form at the CDS via anonymous ftp to cdsarc.u-strasbg.fr (130.79.128.5) or via http://cdsweb.u-strasbg.fr/cgi-bin/qcat?J/A+A/.

\begin{acknowledgements}

This work is based on observations made with the NASA/ESA/CSA James Webb Space Telescope. The data were obtained from the Mikulski Archive for Space Telescopes at the Space Telescope Science Institute, which is operated by the Association of Universities for Research in Astronomy, Inc., under NASA contract NAS 5-03127 for JWST. These observations are associated with program \#1288 (DOI: 10.17909/pg4c-1737).
Support for program \#1288 was provided by NASA through a grant from the Space Telescope Science Institute, which is operated by the Association of Universities for Research in Astronomy, Inc., under NASA contract NAS 5-03127.
Els Peeters and Jan Cami acknowledge support from the University of Western Ontario, the Canadian Space Agency (CSA, 22JWGO1-16), and the Natural Sciences and Engineering Research Council of Canada. 
This article is based upon work from COST Action CA21126 - Carbon molecular nanostructures in space (NanoSpace), supported by COST (European Cooperation in Science and Technology). 
C.B. is grateful for an appointment at NASA Ames Research Center through the San Jos\'e State University Research Foundation (80NSSC22M0107).
T.O. acknowledges the support by the Japan Society for the Promotion of Science (JSPS) KAKENHI Grant Number JP24K07087.

\end{acknowledgements}

\bibliography{90pahfitbib}{}
\bibliographystyle{aa}

\begin{appendix}

\section{AIB sum definitions}
\label{app:sums}

To ensure a consistent use of the fit results produced by PAHFIT and the PDR pack and to allow a direct comparison between different analyses based on the ``total flux'' of certain complexes, it is crucial that the subcomponents are summed consistently.
For this purpose, we provide Table~\ref{tab:sums}, which refers to the wavelengths of the features in the PDR pack.

\begin{table}[bht]
    \centering
    \caption{Definitions of sums of AIB subcomponents defined by PDR pack.}
    \label{tab:sums}
    \begin{tabular}{c|l}
    \hline\hline
    Name & Subcomponents ($\lambda$ in \mum)\\
    \hline
    3.3 & 3.280, 3.295 \\
    3.4 & 3.395, 3.403 \\
    3.4-3.6 ``plateau'' & 3.425, 3.463, 3.516, 3.563\\
    5.2 & 5.238, 5.289\\
    5.7 & 5.632, 5.688, 5.752\\
    6.2 & 6.196, 6.239, 6.341\\
    7.7 & 7.418, 7.552, 7.635, 7.753, 7.823\\
    & 7.854 (broad) \\
    11.2 & 11.196, 11.284, 11.233, 11.381\\
    12.7 & 12.611, 12.727, 12.804\\
    16.4 & 16.402, 16.449\\
    \hline\hline
    \end{tabular}
\end{table}

\section{Tuning details}
\label{app:tuning}

Here, we describe the tuning process for each wavelength segment.
The values for the science pack, chosen in Sects.~\ref{sec:tuning33} through \ref{sec:tuning164} are summarized in Table~\ref{tab:tuning}.
Throughout this discussion, we make extensive use of the detailed figures by \citet{Chown2024} that show a normalized comparison of each profile.

\subsection{The 3.3-3.6~\mum AIB complex}
\label{sec:tuning33}

\begin{figure*}[tb]
\centering
    \includegraphics[scale=0.8]{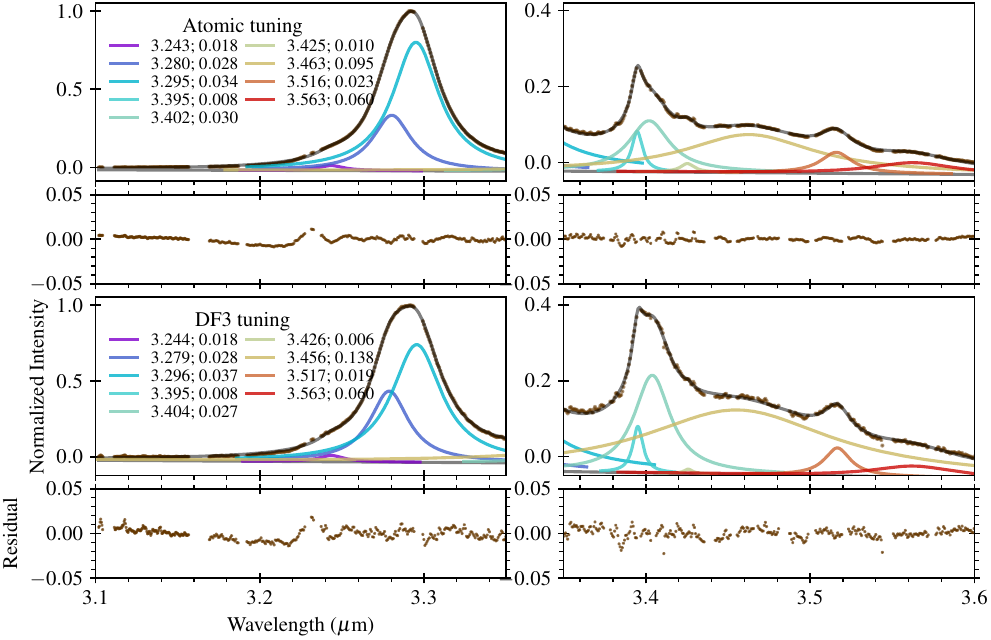}
    \caption{Tuning results for 3.3-3.6~\mum complex based on normalized atomic PDR and DF3 Orion Bar template spectra.
    The rainbow-colored curves show the individual components, and to prevent crowding this visualization, their wings are only shown within three FWHM of their centers.
    The gray curve is their sum, and the brown dots show the normalized spectra used for the fit (see Sect.~\ref{sec:tuning52}).
    The panels at the bottom of each quadrant show the absolute residuals (data - model).
    The numbers in the legend indicate the central wavelengths and FWHM of the individual components.
    Parts of the model can be slightly negative, since a linear term was added to the fits (Sect.~\ref{sec:tuning33}).}
    \label{fig:tuning3336}
\end{figure*}

The bright 3.3 and 3.4~\mum features and the neighboring weaker features from 3.4 to 3.6~\mum were fit as one group, and the results are shown in Fig.~\ref{fig:tuning3336}.
In this figure, the left and right panels show the same fit, split over two panels to present a larger dynamic range.
The selected data range for this group of features is 3.1 to 3.6~\mum, and for the continuum only the linear subtraction method is used.
Since this method uses the edges of the data range as anchor points, it rectifies the spectrum but also brings the flux to zero at these points, which is problematic for fitting the red wings of the features near 3.6~\mum.
Because the Drude profiles have wide wings, fixing the continuum to zero at the edges of the bracketed region will artificially suppress the amplitude and width of some features in the fit result.
To mitigate this, we added a linear function to the tuning model, which allows for flexibility when the wings are non-zero near the edges of the data range.

\citet{Chown2024} show that the 3.3~\mum profile has a slightly variable width, with more variation on the blue side of the profile.
Even with a flexible FWHM, we experience problems with fitting the blue wing if a single component is used: we cannot simultaneously match the shape of the peak and the wings, as the top of the feature does not match the curvature of a single Drude.
Moreover, using a single Drude component makes it impossible to fit the subtle feature near 3.2~\mum; this was already attempted with both Drude and Gaussian profiles by \citet[their appendix G.1]{Peeters2024}, and their Gaussian decomposition worked better for this specific part as they are intrinsically narrower.
The wing of a single Drude profile is already higher than the flux at this wavelength, so no extra flux can be added and the 3.2~\mum amplitude is set to 0.
These problems were mitigated by fitting the 3.3~\mum profile with two narrower components instead of a single broad one, which allows for steeper wings while maintaining the observed width near the peak.

For both the atomic PDR and DF3 spectra, the fitted central wavelengths are near 3.280 and 3.294~\mum, and as shown in Fig.~\ref{fig:tuning3336}, the FWHM are also similar.
As such, two closely spaced components can explain the subtle shape changes of the 3.3~\mum profile by only amplitude changes. 
There is still a slight overestimation of the observed flux in the blue wing, but not strongly so, thereby allowing non-zero amplitudes to be fit to the 3.2~\mum feature.
More focused research comparing many 3.3~\mum profiles in different objects, or as a function of the spatial position in the Orion Bar and other extended objects observed with JWST can clarify how robust this conclusion is.

There is evidence of two components: the atomic PDR template region spectrum exhibits a sharper peak near 3.39~\mum, while the peak has a truncated appearance for the spectrum at DF3.
For the narrow blue-side component of the 3.4~\mum profile we adopt the tuning results of the atomic PDR template region spectrum, since this is where it is most distinct, although the differences between the results for the spectra of the atomic PDR and DF3 template regions are minor.
For the red-side component the FWHM is slightly different and we adopt the average value.
A pair of lines consisting of an H$_2$ line and HI recombination line appears near the peak of the 3.4~\mum profile, and there is a similar pair near the peak center of the red component.
As the high-resolution mode of NIRSpec was used, we were able to precisely mask out the data points corresponding to the lines.
We note that this leaves sufficient data for tuning the 3.4~\mum feature, but for the PAHFIT fits used in the analysis, the lines have to be included in the fit (Sect.~\ref{sec:caveatlines}).

At the base of the red wing of the 3.4~\mum feature, there is a weak feature near 3.42~\mum, for which we adopt the values resulting from fitting the spectrum of the atomic PDR template region, as this is where it is most prominent.
Next is a wide feature with central wavelength observed near 3.46~\mum and wings that extend below the 3.3 to 3.6~\mum features.
The FWHM is rather different between the atomic PDR and DF3 fits.
Given that its width impacts the fit of the 3.42~\mum feature, we adopt the atomic PDR parameters for 3.46~\mum as well.
For the feature near 3.52~\mum, we also adopt the atomic PDR results for the same reason.
Finally, there is a feature that appears to peak near 3.56~\mum.
This feature was difficult to fit, as the results depend strongly on the subtraction of the continuum and the wide structures observed in the 3.8 to 4.5~\mum range, as discussed in Section~\ref{sec:filler}.
The values shown in Fig.~\ref{fig:tuning3336} and Table~\ref{tab:tuning} were obtained after manual adjustment.

\subsection{Approximations for 3.8-4.5~\mum and 6.7-7.0~\mum}
\label{sec:filler}

Before we continue with the discussion of the other complexes, we first note a few regions for which we could not provide a meaningful decomposition, yet the wings of the neighboring complexes are not enough to reproduce the flux.
Instead, Drude profiles were added near these wavelengths to mitigate how this extra flux biases the results for the neighboring complexes.

The first case is the 3.8-4.4~\mum range, where several very broad features seem to be present near 3.8, 4.1, 4.3, and 4.7~\mum, some of which may originate from a vibrational `quasi continuum' of PAHs \citep{Allamandola1989, Boersma2023}.
This region is complicated further by the NIRSpec chip gap at 3.98-4.17 \mum and the presence of CO and other molecular lines  \citep{Peeters2024}.
There could also be continuum contributions that cannot be reproduced by modified blackbody curves (Sect.~\ref{sec:caveatcontinuum}).
Further difficulties in this wavelength range include potential absorption by CO$_2$ ice near 4.27~\mum, and ambiguity for the reduction in flux at 4.50~\mum, which could either be absorption or simply a decrease of flux between two emission features.
For a detailed overview of this wavelength range, we refer to \citet[Figs. 6~and B.7]{Peeters2024}.
In the science pack, we have introduced four wide features, with the approximate manually set parameters given in Table~\ref{tab:tuning}.
Adding these features prevents the fits of the 3.6 and 5.2~\mum features from being strongly biased by the flux in this region.
To find a tuning that can properly fit some of these feature, in particular the deuterated aromatic and aliphatic features \citep{Hudgins2004, Buragohain2015, Allamandola2021, Yang2023}, a different set of spectra will be needed.

The 6.8-7.0~\mum data points were omitted due to artifacts resembling peaks and troughs near the [Ar II] 6.99~\mum line \citep[see][]{VanDePutte2024}, introduced by an overcorrection of the stray light caused by this bright line.
To approximate the remaining data points between the 6.2~\mum and 7.7~\mum complexes, similar wide and approximate features were added near 6.7 and 7.0~\mum.
This helps with fitting the  red wing of the 6.2~\mum feature and the blue wing of the 7.7~\mum feature.

Summarizing, despite the detailed tuning process and various methods to deal with the continuum, there is some ambiguity near the ends of each complex for which we provide a tuning, considering that the contents of the regions discussed above are unclear.
The physicality of these additions should become more clear when their spatial distributions are compared in a variety of objects, which we defer to future work.

\subsection{The 5.2 and 5.7~\mum complexes}
\label{sec:tuning52}

\begin{figure*}[tb]
\centering
\includegraphics[scale=0.8]{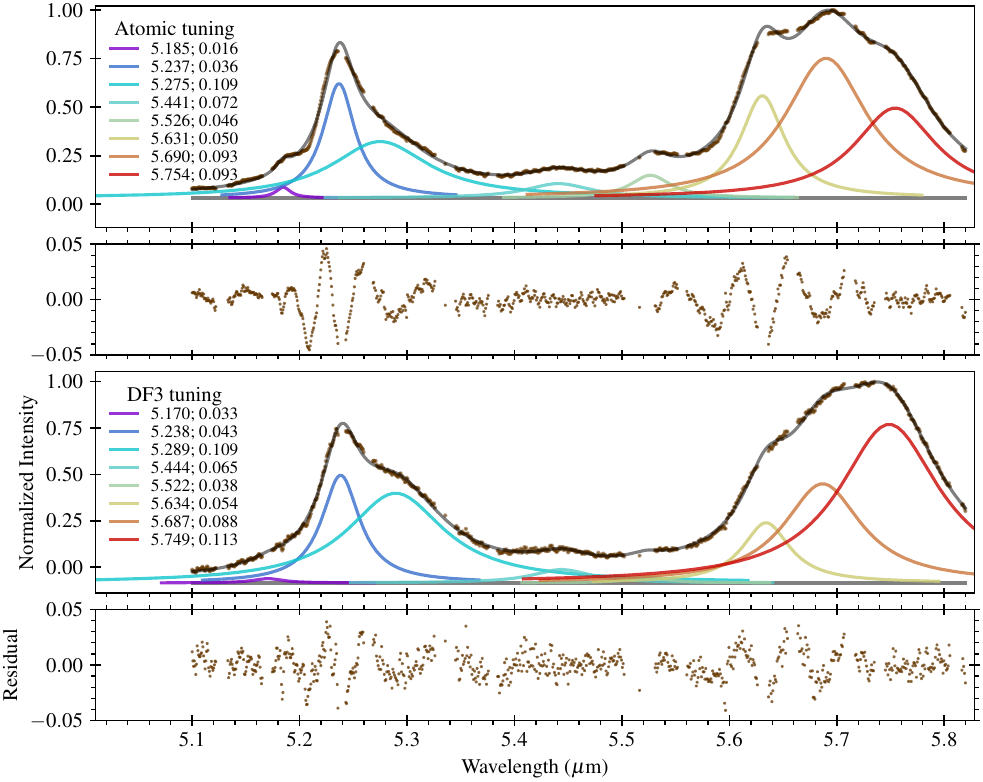}
\caption{Tuning results for the 5.2 and 5.7~\mum complexes. 
(See Figure~\ref{fig:tuning3336} for details.)}
\label{fig:tuning52}
\end{figure*}

Since the 5.2~\mum complex is the weakest one we are tuning, with the lowest S/N, the wings of the other features will have a significant impact on it.
For example, contributions from the blue wing of the bright 6.2~\mum profile and the feature near 5.87~\mum extend into the 5.2-5.7~\mum range.
Therefore, subtracting the wings of the 6.2~\mum feature was crucial here.
The cleaned and rectified spectra and the fit applied to them to derive the tuning are shown in Fig.~\ref{fig:tuning52}.

Our decomposition for the 5.2~\mum profile consists of one narrow peak located at the observed maximum near 5.24~\mum, a second broader component in the red wing, which is very prominent in the DF3 spectrum, and a small component near 5.18~\mum at the base of the blue wing, which only seems to appear in the spectrum of the atomic PDR template region.
For the tuning, the wavelength and FWHM are based on the atomic PDR template region spectrum for 5.18~\mum, and on that from DF3 for 5.24 and 5.29~\mum.

The complex near 5.7~\mum consists of three features for the atomic PDR spectrum, and their ratios are different for DF3.
Between the atomic PDR and the DF3 spectrum, the central wavelengths and FWHMs only differ slightly ($\lesssim 0.004$), so we adopt the mean values and keep the parameters fixed.

Between the 5.2 and 5.7~\mum complexes, the atomic PDR spectrum has two oscillations with local maxima near 5.45 and 5.53~\mum.
We base the tuning for these two features on the atomic PDR spectrum, and confirmed that these settings also work for the local maximum in the DF3 spectrum near 5.45~\mum, though the 5.53~\mum feature is not significantly present for DF3.

\subsection{The 6.2~\mum complex}
\label{sec:tuning62}

\begin{figure}[tbh]
\centering
    \includegraphics[scale=0.8]{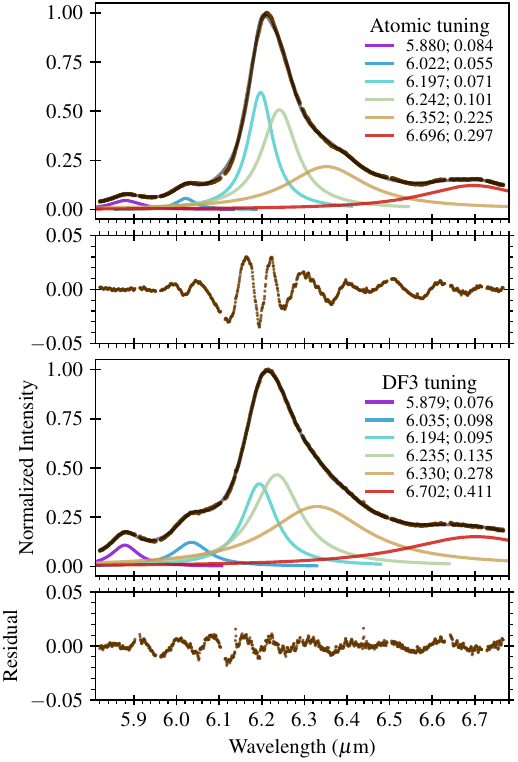}
    \caption{Tuning of 5.9, 6.0, and 6.2~\mum features and approximation for tentative features near 6.7~\mum.}
    \label{fig:tuning62}
\end{figure}

To prepare a wavelength segment near 6.2~\mum for the tuning, we needed to subtract the contributions by filler features described in Sect.~\ref{sec:filler}.
As for the other complexes, this was done using an intermediate PAHFIT run.
A linear continuum correction was not applied. 
The resulting segment and tuning fit are shown in Fig.~\ref{fig:tuning62}.

The 6.2~\mum profile has a strong asymmetry, with a steep blue wing and an extended red wing.
The differences seen near 6.3~\mum indicate a widening of the red wing for DF2 and DF3 compared to the atomic PDR spectrum. 
This red wing is also less concave for the DF templates.
The peak position and the width of the blue wing seems mostly constant, with only a slight apparent change in width.
To reproduce both the asymmetry and the changes in width, we use two features to fit the main peak at 6.2~\mum. 
We could not reproduce the 6.2~\mum profile for both the atomic PDR and DF3 spectra with fixed width components and therefore allowed the FWHM to vary.
We find that in practice (Sect.~\ref{sec:tuningvalidation}), the range for the FWHM needs to be somewhat larger than just the range between the atomic PDR and DF3 values, hence the larger interval provided in Table~\ref{tab:tuning}.

Near 6.4~\mum, there is a feature present in the atomic PDR spectrum, while the wing looks smoother in DF3, so we adopt the values for the FHWM and wavelength based on the fit to the atomic PDR spectrum.
In the tuning presented here, the FWHM is actually too wide to accurately reproduce the 6.4~\mum feature, as the feature we introduced also accounts for some of the contributions by a weaker feature near 6.5~\mum.
Between 6.6~\mum and 6.7~\mum, there exist two more local maxima which we approximate by a single wide component for which the mean values are adopted.
For optimal results, the 6.2 and 7.7~\mum complexes may need to be tuned together in an all-wavelength fit. 
The main caveat for the weaker features on the red side of the 6.2~\mum complex is that the results depend on the assumptions for the 6.4-7.0~\mum range (see Sect.~\ref{sec:filler}).

On the blue side, the features in the 5.8-6.1~\mum range are most pronounced for DF3, so we adopt the results from this fit into our tuning.
Having a flexible FWHM for the 6.2~\mum components is important for fitting these weaker features, as the wings of the strong 6.2~\mum feature need to be reasonably reproduced.
If this is not the case, and the assumed FWHM for the 6.2~\mum profile is too broad, then the blue wing of the latter will exceed the total flux in the data, leaving no room for the 6.0~\mum feature.
As a caveat, occasional unreliable fits of the 5.8 and 6.0~\mum amplitude can be expected.

\subsection{The 7.7 and 8.6~\mum complexes}
\label{sec:tuning7786}

\begin{figure*}[tb]
    \centering
    \includegraphics[scale=0.8]{{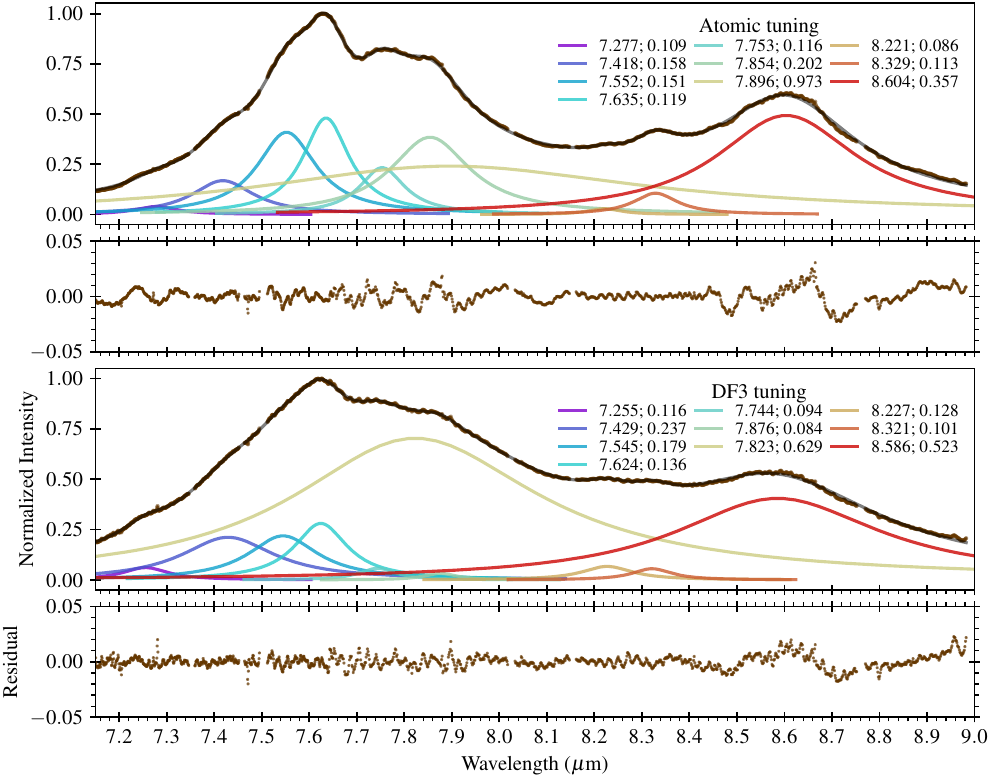}}
    \caption{Tuning of the 7.7 and 8.6~\mum complexes.}
    \label{fig:tuning7786}
\end{figure*}

For the 7.7 and 8.6~\mum complexes, we used the data from 7.15 to 9.1~\mum and apply the PAHFIT-based continuum subtraction, but not the residual linear continuum subtraction.

The 7.7 complex profile has three local maxima, which are most pronounced in the atomic PDR spectrum.
This indicates at least three components, for which the profiles are relatively narrow compared to the total width of the complex.
Additional substructure is seen in the blue wing.
There is evidence of an additional component near 7.55~\mum, on the blue side of the 7.6~\mum peak.
In the atomic PDR spectrum, this additional feature appears as a subtle transition from convex to concave near 7.59~\mum.
A similar blue shoulder feature appears in the atomic PDR spectrum near 7.42~\mum.
Finally, a feature near the base of the 7.7~\mum blue wing is seen at approximately 7.25~\mum, which is strongest in DF2 and DF3.
Therefore, we set up six components for the blue side and the peak of the 7.7~\mum profile.

An additional detail is that the depth of the gap between the 7.60 and 7.75~\mum features appears to change, together with the general slope near 7.8~\mum. A similar change in slope is seen from 8.25 to 8.6~\mum. 
The blue wing of 7.7~\mum is also smoother for the DF3 spectrum, meaning that the features cannot be as pronounced, but the total flux still has to add up to the observed value.
We hypothesize that these effects are the result of a wide feature somewhere in the range of 7.5 to 8.0~\mum.
For the atomic PDR spectrum, the contribution of this wide component is weaker, resulting in a more pronounced appearance of the 7.7~\mum profile details described above.
Given that the red 7.7~\mum wing is concave for the atomic PDR spectrum, while it resembles a straight line segment for the DF3 spectrum, we place the central wavelength near 7.8~\mum as an initial guess.

In total, it appears the 7.7~\mum profile needs to be described by seven components, and we have fit them in Fig.~\ref{fig:tuning7786}.
For the features corresponding to the three maxima of the profile and the details of the blue wing, we adopt the values obtained from the atomic PDR spectrum, except for the 7.25~\mum feature, which is more pronounced in DF3.
We note that the best-fit FWHM values for DF3 are significantly wider for the features of the blue wing.
This may mean that the addition of the 7.8~\mum feature is not enough to allow for a smooth fit of the blue wing, and perhaps there could be a different solution.
For the wide feature, the best fit wavelength and FWHM are different between the atomic PDR and DF3 results and we adopt the DF3 values with constant FWHM since this is where the fit results in the largest flux contribution by the wide component.

For the profile that peaks near 8.6~\mum, the shape and the change in slope at the peak of the profile indicates that there may be two components.
Another indication is that the profile is slightly concave for DF1 and DF2.
However, the change in slope could also be partially explained as an effect of the wing of the wide 7.9~\mum feature. 
Despite this detail, the 8.6~\mum profile seems well reproduced by a single component and adding an additional component near for example 8.55~\mum does not seem to improve the fit as the fitter sets the amplitude of such additions to zero.
We continue with a single component for the remainder of this work, and use the atomic PDR tuning as the influence of the broad 7.9~\mum component is smaller there.

Between the 7.7 and 8.6~\mum peaks, there are two weak features near 8.2 and 8.3~\mum, and the latter is stronger in the atomic PDR spectrum.
We adopt the values from the DF3 fit for 8.2~\mum, and that from the atomic PDR fit for 8.3~\mum.
We note that \citet{Chown2024} identified additional features near 7.05 and 8.78~\mum. These are weak and we do not include them for now.

\subsection{The 11.2 and 12.7~\mum complexes}
\label{sec:tuning1014}

\begin{figure*}[tb]
    \centering
    \includegraphics[scale=0.8]{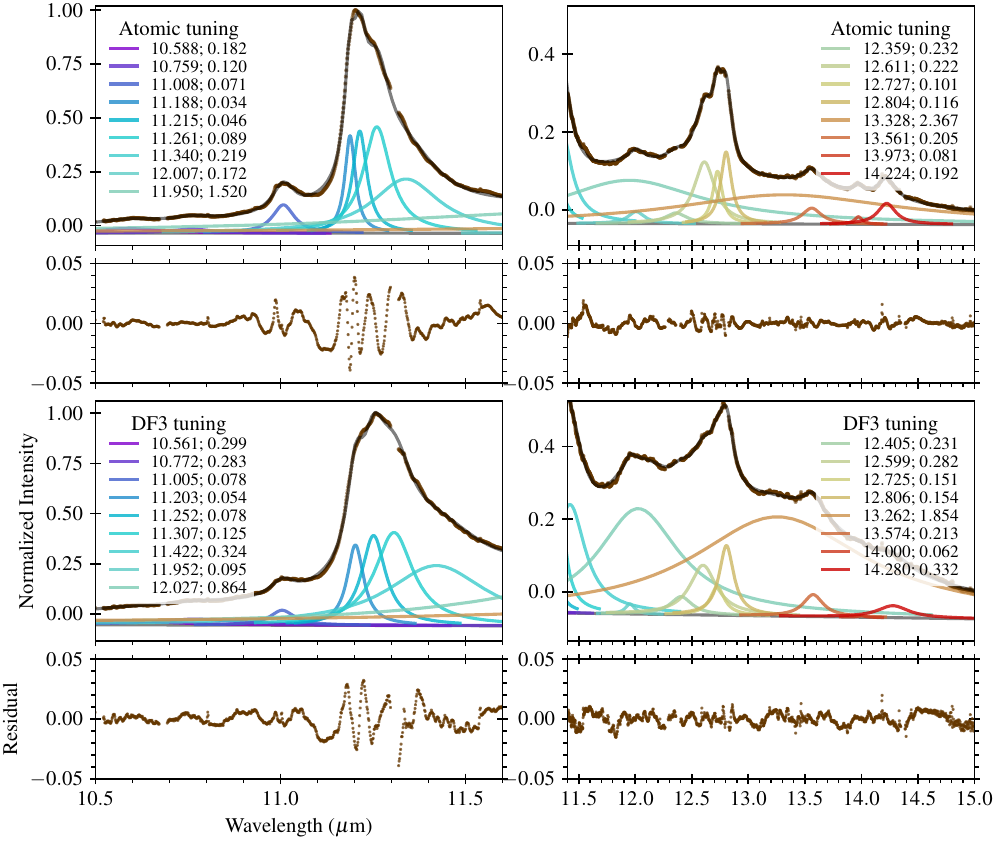}
    \caption{Tuning of features in the 10 to 15~\mum wavelength range.}
    \label{fig:tuning1014}
\end{figure*}

With the shape of the continuum we assumed in Fig. \ref{fig:continuum}, the features between 10.5 and 14~\mum all overlap and form one complex.
A caveat is that we introduced two wide features for which the optimal tuning depends on the choice of the continuum.
The strongest of these features is the wide feature at 12.00~\mum, as evidenced by the rise in flux between 11.8 and 12.2~\mum, which is most prominent in DF3.
A similar but more subtle effect seen is near 13.3~\mum. 
The fits shown in Fig.~\ref{fig:tuning1014} therefore include a wide component near 12~\mum and one near 13.3~\mum. 
The curvature of the peak of these components is noticeable in the DF3 data but harder to differentiate in the atomic PDR data, resulting in very wide profiles for the latter.
Just as for the 7.7~\mum profile, the wide components seem to have a much stronger relative contribution in the DF3 spectrum and we adopt the DF3 results for which the FWHM values are significantly narrower.

We note that there is some substructure at the peak of the broad 12.0 \mum feature, which is not well fitted by a single wide profile; a narrower feature was introduced on top of the 12~\mum component to approximate this substructure.
A minor caveat is the known spectral leak in the MRS data resulting in a flux artifact that appears in the 12.0 to 12.4~\mum range and peaks near 12.2~\mum, and we have not corrected for this.

Multiple components are needed to fit each of the two main peaks at 11.2 and 12.7~\mum.
The position of the maximum is shifted when compared between the atomic PDR and DF spectra.
Moreover, the atomic PDR and DF1 spectra show signs of a feature that peaks at the same positions as DF2 and DF3, indicating that at least two components are present.
In total we introduced four components, of which three were placed between 11.20 and 11.29~\mum as an initial guess.
A third component was added in between the two observed positions of the maximum because the DF2 profile also appears to have a feature on its blue side, of which the apparent wavelength is slightly redder than the atomic PDR peak.
We note that the exact positions of the narrower features are harder to identify in this wavelength range because channel 2 LONG of MIRI MRS is particularly susceptible to residual fringes in the data.
A more advanced residual fringe correction may make it more straightforward to identify if there are two or three components at 11.2~\mum.
The fourth component is there to reproduce the extended red wing.
We adopt the mean FWHM and central wavelengths for these four 11.2~\mum components in the science pack.

For the 12.7~\mum profile we use two components since the peak has a flat top in the atomic PDR spectrum while the DF2 and DF3 spectra peak on the red side. 
Inspecting the blue wing of the 12.7~\mum profile, there is another clear feature near 12.6~\mum and a more subtle one near 12.35~\mum.
Since these four features are most distinct in the atomic PDR spectrum, we adopt those tuning results in the science pack.

Red- and blue-ward of the 11.2 to 12.7~\mum complex there are several additional features, all of which are most prominent in the atomic PDR spectrum, 
while some are very weak in DF3.
For all the following features in this paragraph the atomic PDR tuning results were therefore adopted as the DF3 results can be unreliable.
For the weak features near 10.5 and 10.8~\mum is a caveat that the tuning results depend on  the 10 \mum silicate feature of the continuum model.
For the 11.0~\mum feature the results also depend on how the fit of the 11.2~\mum profile is configured.
Redward of 12.7~\mum we find three more features at 13.5, 14, and 14.2~\mum, which are fit with single components.

\subsection{The 16.4 and 17.4~\mum complexes}
\label{sec:tuning164}

\begin{figure*}[tb]
    \centering
    \includegraphics[scale=0.8]{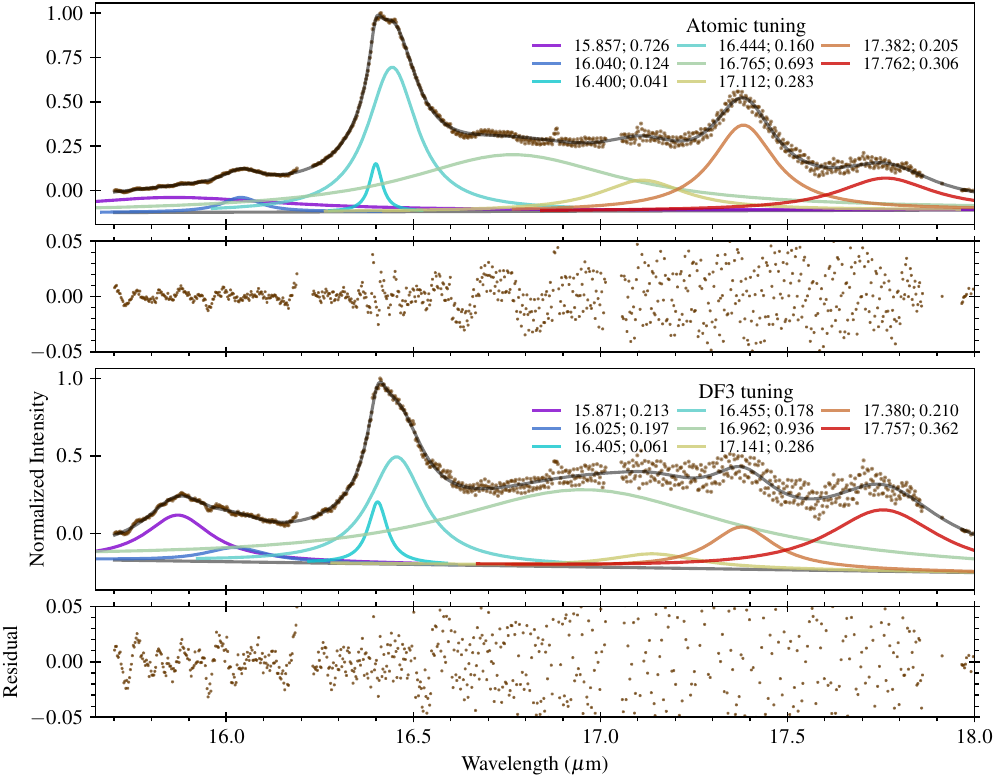}
    \caption{Tuning results for the 16 to 18~\mum range.}
    \label{fig:tuning1618}
\end{figure*}

For the 16 to 18~\mum wavelength range we subtracted the continuum using both the preliminary PAHFIT model and the linear subtraction of Sect.~\ref{sec:tuningstrategy}. 
The latter was done because the alternate continuum model (Sect.~\ref{sec:pahfitcontinuummodel}) underestimates the intensity near 17.7~\mum somewhat. The extra linear term was also added to the model and the results are shown in Fig.~\ref{fig:tuning1618}.

The feature near 15.8~\mum is prominently observed in the DF3 spectrum, while it looks as though it is a small extension of another feature in the atomic PDR.
For the feature at 16.0~\mum the situation is opposite, so we adopt the atomic PDR results for 15.8~\mum and the DF3 results for 16.0~\mum.

The 16.4~\mum profile exhibits a subtle double peak nature for the atomic PDR spectrum, with the second peak at 16.45~\mum, while the DF3 spectrum is more convex around the same wavelength.
We adopt the mean tuning values for these two features, keeping them fixed.
For the 17.4~\mum feature only one component is identifiable at the S/N of these data.
We adopt a single feature with the atomic PDR tuning since the S/N is higher there.
Finally, we do the same for the feature near 17.7~\mum.

In between the 16.4 and 17.4~\mum features we model the  spectrum with two wide features, with their extended wings providing significant intensity contributions that overlap with the other features.
There are also two small bumps near 16.7~\mum, providing evidence of real features being present there. 
We introduce wide features at 16.7 and 17.1~\mum, serving as an approximation for the flux between 16.4 and 17.4~\mum.
We adopted the atomic PDR tuning results.

\onecolumn
\begin{longtable}{cccccclc}
\caption{Tuning results (\mum units) and selected values for PDR pack.}
\label{tab:tuning}\\
\hline\hline
Atomic PDR && DF3 && Mean && Bounds & Selected values \\
$\lambda$ & FWHM & $\lambda$ & FWHM & $\lambda$ & FWHM & FWHM & \\ 
\hline
\endfirsthead
\caption{continued.}\\
\hline\hline
Atomic PDR    &      & DF3       &      & Mean      &      & Bounds & Selected values \\
$\lambda$ & FWHM & $\lambda$ & FWHM & $\lambda$ & FWHM & FWHM & \\ 
\hline
\endhead
\endfoot

\multicolumn{7}{c}{\textbf{3.3-3.6} (Sect.~\ref{sec:tuning33})} \\
3.243 & 0.018 & 3.244 & 0.018 & 3.243 & 0.018 & & Mean \\
3.280 & 0.028 & 3.279 & 0.028 & 3.280 & 0.028 & & Mean \\
3.295 & 0.034 & 3.296 & 0.037 & 3.295 & 0.035 & & Mean \\
3.395 & 0.008 & 3.395 & 0.008 & 3.395 & 0.008 & & Atomic PDR \\
3.402 & 0.030 & 3.404 & 0.027 & 3.403 & 0.028 & & Mean \\
3.425 & 0.010 & 3.426 & 0.006 & 3.426 & 0.008 & & Atomic PDR \\
3.463 & 0.095 & 3.456 & 0.138 & 3.459 & 0.116 & & Atomic PDR \\
3.516 & 0.023 & 3.517 & 0.019 & 3.516 & 0.021 & & Atomic PDR \\
3.563 & 0.060 & 3.563 & 0.060 & 3.563 & 0.039 & & Manual \\

\multicolumn{7}{c}{\textbf{Approximation 3.8-4.8} (Sect.~\ref{sec:filler})} \\
-     & -     & -     & -     & 3.795 & 0.323 & & Manual \\
-     & -     & -     & -     & 4.077 & 0.232 & & Manual \\
-     & -     & -     & -     & 4.347 & 0.303 & & Manual \\
-     & -     & -     & -     & 4.758 & 0.955 & & Manual \\

\multicolumn{7}{c}{\textbf{5.2 and 5.7} 
(Sect.~\ref{sec:tuning52})} \\
5.185 & 0.016 & 5.170 & 0.033 & 5.178 & 0.025 & & Atomic PDR \\
5.237 & 0.036 & 5.238 & 0.043 & 5.237 & 0.040 & & DF3  \\
5.275 & 0.109 & 5.289 & 0.109 & 5.282 & 0.109 & & DF3  \\
5.441 & 0.072 & 5.444 & 0.065 & 5.442 & 0.069 & & Atomic PDR  \\
5.526 & 0.046 & 5.522 & 0.038 & 5.524 & 0.042 & & Atomic PDR  \\
5.631 & 0.050 & 5.634 & 0.054 & 5.632 & 0.052 & & Mean  \\
5.690 & 0.093 & 5.687 & 0.088 & 5.688 & 0.090 & & Mean  \\
5.754 & 0.093 & 5.749 & 0.113 & 5.752 & 0.103 & & Mean  \\

\multicolumn{7}{c}{\textbf{6.2} (Sect.~\ref{sec:tuning62})} \\
5.880 & 0.084 & 5.879 & 0.076 & 5.880 & 0.080 & & Mean  \\
6.022 & 0.055 & 6.035 & 0.098 & 6.028 & 0.077 & & Mean  \\
6.197 & 0.071 & 6.194 & 0.095 & 6.196 & 0.083 & (0.058, 0.100) & Mean  \\
6.242 & 0.101 & 6.235 & 0.135 & 6.239 & 0.118 & (0.100, 0.160) & Mean  \\
6.352 & 0.225 & 6.330 & 0.278 & 6.341 & 0.252 & & Mean  \\

\multicolumn{7}{c}{\textbf{Approximation 6.6-7.0} (Sects.~\ref{sec:filler} and~\ref{sec:tuning62})} \\
6.696 & 0.297 & 6.702 & 0.411 & 6.699 & 0.354 & & Mean  \\
-     & -     & -     & -     & 7.030 & 0.468 & & Manual \\

\multicolumn{7}{c}{\textbf{7.7 and 8.6} (Sect.~\ref{sec:tuning7786})} \\
7.277 & 0.109 & 7.255 & 0.116 & 7.266 & 0.112 & & DF3  \\
7.418 & 0.158 & 7.429 & 0.237 & 7.424 & 0.197 & & Atomic PDR  \\
7.552 & 0.151 & 7.545 & 0.179 & 7.549 & 0.165 & & Atomic PDR  \\
7.635 & 0.119 & 7.624 & 0.136 & 7.630 & 0.127 & & Atomic PDR  \\
7.753 & 0.116 & 7.744 & 0.094 & 7.749 & 0.105 & & Atomic PDR  \\
7.854 & 0.202 & 7.876 & 0.084 & 7.865 & 0.143 & & Atomic PDR  \\
7.896 & 0.973 & 7.823 & 0.629 & 7.860 & 0.801 & & DF3  \\
8.221 & 0.086 & 8.227 & 0.128 & 8.224 & 0.107 & & DF3  \\
8.329 & 0.113 & 8.321 & 0.101 & 8.325 & 0.107 & & Atomic PDR  \\
8.604 & 0.357 & 8.586 & 0.523 & 8.595 & 0.440 & & Atomic PDR \\

\multicolumn{7}{c}{\textbf{11.2 and 12.7} (Sect.~\ref{sec:tuning1014})} \\
10.588 & 0.182 & 10.561 & 0.299 & 10.574 & 0.241 & & Atomic PDR  \\
10.759 & 0.120 & 10.772 & 0.283 & 10.765 & 0.202 & & Atomic PDR  \\
11.008 & 0.071 & 11.005 & 0.078 & 11.006 & 0.075 & & Atomic PDR  \\
11.188 & 0.034 & 11.203 & 0.054 & 11.196 & 0.044 & & Mean  \\  
11.215 & 0.046 & 11.252 & 0.078 & 11.233 & 0.062 & & Mean  \\  
11.261 & 0.089 & 11.307 & 0.125 & 11.284 & 0.107 & & Mean  \\  
11.340 & 0.219 & 11.422 & 0.324 & 11.381 & 0.271 & & Mean  \\  
12.007 & 0.172 & 11.952 & 0.095 & 11.980 & 0.133 & & DF3  \\
11.950 & 1.520 & 12.027 & 0.864 & 11.989 & 1.192 & & DF3  \\   
12.359 & 0.232 & 12.405 & 0.231 & 12.382 & 0.231 & & Atomic PDR  \\
12.611 & 0.222 & 12.599 & 0.282 & 12.605 & 0.252 & & Atomic PDR  \\
12.727 & 0.101 & 12.725 & 0.151 & 12.726 & 0.126 & & Atomic PDR  \\
12.804 & 0.116 & 12.806 & 0.154 & 12.805 & 0.135 & & Atomic PDR  \\
13.328 & 2.367 & 13.262 & 1.854 & 13.295 & 2.110 & & DF3  \\   
13.561 & 0.205 & 13.574 & 0.213 & 13.567 & 0.209 & & Atomic PDR  \\
13.973 & 0.081 & 14.000 & 0.062 & 13.987 & 0.072 & & Atomic PDR  \\
14.224 & 0.192 & 14.280 & 0.332 & 14.252 & 0.262 & & Atomic PDR  \\

\multicolumn{7}{c}{\textbf{16.4 and 17.4} (Sect.~\ref{sec:tuning164})} \\
15.857 & 0.726 & 15.871 & 0.213 & 15.864 & 0.469 & & DF3 \\
16.040 & 0.124 & 16.025 & 0.197 & 16.032 & 0.161 & & Atomic PDR \\
16.400 & 0.041 & 16.405 & 0.061 & 16.402 & 0.051 & & Mean \\
16.444 & 0.160 & 16.455 & 0.178 & 16.449 & 0.169 & & Mean \\
16.765 & 0.693 & 16.962 & 0.936 & 16.863 & 0.815 & & Atomic PDR \\
17.112 & 0.283 & 17.141 & 0.286 & 17.127 & 0.284 & & Atomic PDR \\
17.382 & 0.205 & 17.380 & 0.210 & 17.381 & 0.208 & & Atomic PDR \\
17.762 & 0.306 & 17.757 & 0.362 & 17.760 & 0.334 & & Atomic PDR \\
\hline                                             
\hline
\end{longtable}
\tablefoot{If no bounds are listed, this means the FWHM is fixed.}

\twocolumn

\end{appendix}

\end{document}